\documentclass[a4paper,11pt]{article}

\usepackage{jcappub}
\usepackage{graphicx}
\usepackage{bm}
\usepackage{latexsym}
\usepackage{amsfonts,amssymb}
\usepackage{amsmath}
\usepackage{color}

\makeatletter
\gdef\@fpheader{}
\makeatother

\def\refs{\marginpar {\scriptsize REFERENCES}}

\def\beq{\begin{equation}}
\def\eeq{\end{equation}} 
\def\bea{\begin{eqnarray}}
\def\eea{\end{eqnarray}}
\def\benu{\begin{enumerate}}
\def\eenu{\end{enumerate}}
\def\nn{\nonumber} 

\def\f{\frac}
\def\l{\left}
\def\r{\right}
\def\d{{\rm d}}

\def\ei{\eta_{\rm i}}
\def\ee{\eta_{\rm e}}

\def\vk{{\bm k}}

\def\cG{{\cal G}}
\def\fnl{f_{_{\rm NL}}}
\def\Mpl{M_{_{\rm Pl}}}
\def\ns{n_{_{\rm S}}}

\newcommand{\uPl}{\mathrm{Pl}}

\newcommand{\usssPl}{\sss{\uPl}}

\newcommand{\Mp}{M_\usssPl}

\newcommand{\sss}[1]{{\scriptscriptstyle{#1}}}

\subheader{Sharp inflaton potentials and bi-spectra}
\title{Sharp inflaton potentials and bi-spectra:~Effects 
of smoothening the discontinuity}
\author[a]{J\'er\^ome Martin,}
\author[b]{L.~Sriramkumar} 
\author[c]{and Dhiraj Kumar Hazra}

\affiliation[a]{Institut d'Astrophysique de Paris, UMR7095-CNRS,
  Universit\'e Pierre et Marie Curie, 98bis boulevard Arago, 75014
  Paris, France.}

\affiliation[b]{Department of Physics, Indian Institute of Technology
  Madras, Chennai~600036, India.}

\affiliation[c]{Asia Pacific Center for Theoretical Physics, Pohang,
  Gyeongbuk 790-784, Korea.}

\emailAdd{jmartin@iap.fr}
\emailAdd{sriram@physics.iitm.ac.in}
\emailAdd{dhiraj@apctp.org}
\date{today} 

\begin{document}

\abstract{Sharp shapes in the inflaton potentials often lead to short
  departures from slow roll which, in turn, result in deviations from
  scale invariance in the scalar power spectrum.  Typically, in such
  situations, the scalar power spectrum exhibits a burst of features
  associated with modes that leave the Hubble radius either immediately
  before or during the epoch of fast roll.  Moreover, one also finds
  that the power spectrum turns scale invariant at smaller scales
  corresponding to modes that leave the Hubble radius at later stages,
  when slow roll has been restored.  In other words, the imprints of
  brief departures from slow roll, arising out of sharp shapes in the
  inflaton potential, are usually of a finite width in the scalar
  power spectrum.  Intuitively, one may imagine that the scalar
  bi-spectrum too may exhibit a similar behavior, i.e. a restoration of
  scale invariance at small scales, when slow roll has been
  reestablished.  However, in the case of the Starobinsky model 
  (viz.  the model described by a linear inflaton potential with a sudden 
  change in its slope) involving the canonical scalar field, it has been 
  found that, a rather sharp, though
  short, departure from slow roll can leave a lasting and significant
  imprint on the bi-spectrum.  The bi-spectrum in this case is found
  to grow linearly with the wavenumber at small scales, a behavior 
  which is clearly unphysical. In this work, we study the effects of
  smoothening the discontinuity in the Starobinsky model on the scalar
  bi-spectrum. Focusing on the equilateral limit, we 
  analytically show that, for smoother potentials, the bi-spectrum indeed 
  turns scale invariant at suitably large wavenumbers.  We also confirm 
  the analytical results numerically using our newly developed code BINGO. 
  We conclude with a few comments on certain related points.}


\maketitle


\section{Inflationary models, discontinuities and the scalar bi-spectrum}

The inflationary scenario is a very efficient paradigm to resolve the
puzzles of the standard cosmological model and to simultaneously describe 
the origin of perturbations in the early
universe~\cite{Kolb:1990aa,Dodelson:2003ft,Mukhanov:2005sc,Weinberg:2008zzc,Durrer:1127831,Lyth:2009zz,PPJPU,mo2010,Lemoine:2008zz,Kodama:1985bj,Mukhanov:1990me,Lidsey:1995np,Lyth:1998xn,Riotto:2002yw,Martin:2007bw,Martin:2004um,Martin:2003bt,Bassett:2005xm,Kinney:2009vz,Sriramkumar:2009kg,Baumann:2009ds}. 
Even the simplest of models lead to a sufficiently long duration of 
inflation that is required to overcome the horizon problem. Moreover, 
many of these models permit inflation of the slow roll type, which 
generates a nearly scale invariant primordial power spectrum that is 
remarkably consistent with the observations of the anisotropies in the 
Cosmic Microwave Background (CMB) and other cosmological
data~\cite{Larson:2010gs,Komatsu:2010fb,Bennett:2012fp,Hinshaw:2012fq,Ade:2013lta,Planck:2013kta,Ade:2013ydc}.

\par

While attempting to identify the correct inflationary scenario, apart 
from the power spectrum, the non-Gaussianities and, in particular, the 
scalar bi-spectrum, also play a significant role. Indeed, the recent Planck 
data have shown that the non-Gaussianities are consistent with zero, with 
the three parameters that characterize the scalar bi-spectrum constrained 
to be: $\fnl^{\rm loc}=2.7\pm 5.8$, $\fnl^{\rm eq}=-42\pm 75$ and $
\fnl^{\rm ortho}=-25\pm 39$ \cite{Ade:2013ydc}. 
These constraints imply that the correct model of inflation cannot deviate 
too much from the standard single field inflation of the slow roll type,
involving the canonical kinetic term.
In other words, inflation seems to be a non-trivial (i.e. $\ns \neq 1$, 
where $\ns$ denotes the scalar spectral index), but `non-exotic' (viz.
$\fnl \simeq 0$) mechanism~\cite{Martin:2010hh,Martin:2013nzq,Martin:2013tda}. 
On the theoretical front, the most complete formalism to calculate the 
three-point correlation functions involving scalars and tensors in a given
inflationary model is the approach due to Maldacena~\cite{Maldacena:2002vr}.  
In the Maldacena formalism, the three-point functions are evaluated using the 
standard rules of perturbative quantum field theory, based on the interaction
Hamiltonian that depends cubically on the
perturbations~\cite{Maldacena:2002vr,Seery:2005wm,Chen:2005fe,Chen:2006nt,Langlois:2008qf,Langlois:2008wt,Chen:2010xka}.
The resulting expressions for the three-point function of primary
interest, viz. the scalar bi-spectrum, involves integrals over
combinations of the background quantities such as the scale factor and
the slow roll parameters as well as the modes describing the curvature
perturbation (see, for instance,
Refs.~\cite{Martin:2011sn,Hazra:2012kq}).  Evaluating the bi-spectrum
analytically for a generic inflationary model proves to be a
non-trivial task.  But, as in the case of the power spectrum, the
bi-spectrum can be calculated analytically under the slow roll
approximation~\cite{Maldacena:2002vr,Seery:2005wm,Chen:2006nt,Gangui:1993tt,Gangui:1994yr,Gangui:1999vg,Wang:1999vf}.

\par

As we have already mentioned, slow roll inflation driven by a single scalar 
field seems to be the most likely possibility to describe the early universe. 
Nevertheless, it is also interesting to consider other scenarios which 
lead to larger levels of non-Gaussianities.
Such analyses can help us gain a better understanding of the constraints 
imposed by the Planck data on the parameters characterizing these class 
of models. 
Moreover, these exercises can actually allow us to assess the degree of 
fine-tuning implied by the CMB data from Planck and WMAP on the non-minimal alternatives
(see, for example, Ref.~\cite{Hazra:2013xva,Hazra:2013nca,Dorn:2014kga}). 
Further, the recent claim of the detection of the imprints of the primordial 
tensor modes by BICEP2 and the indication of a relatively high tensor-to-scalar 
ratio~\cite{Ade:2014xna,Ade:2014gua}, if confirmed, implies that we cannot 
completely rule out non-trivial possibilities either.
Studying non-standard scenarios is however not a simple task, since, in 
these situations, the calculation of non-Gaussianities can be highly 
non-trivial and one often has to rely on numerical calculations (for 
numerical analysis of specific models, 
see Refs.~\cite{Chen:2006xjb,Chen:2008wn,Hotchkiss:2009pj,Hannestad:2009yx,Flauger:2010ja,Adshead:2011bw,Adshead:2011jq};
for a broader discussion on the procedures involved and applications
to a few different classes of models, see Ref.~\cite{Hazra:2012yn};
in this context, also see Ref.~\cite{Sreenath:2013xra}). 
However, occasionally, it is also possible to evaluate the bi-spectrum 
analytically in non-trivial situations such as scenarios involving 
departures from slow roll~\cite{Starobinsky:1992ts,Dvorkin:2009ne,Hu:2011vr}. 
One such example that permits an analytic evaluation of the power spectrum 
and the complete bi-spectrum (at least, in the equilateral limit) even in
the presence of fast roll, is the model originally due to
Starobinsky~\cite{Starobinsky:1992ts}. This model has recently
attracted quite a lot of attention, but different physical conclusions
with regards to the shape of the bi-spectrum have been reached. The
present paper is aimed at considering the question again in order to
clarify the situation.

\par

As we shall soon outline, the Starobinsky model involves a canonical scalar 
field and is described by a linear potential with a sudden change in the
slope at a given point.  The sharp change in the slope leads to a
brief period of fast roll sandwiched between two epochs of slow roll.
Typically, in such situations, the power spectrum is expected to turn
scale invariant when slow roll has been restored, and it is indeed
what happens in the case of the Starobinsky model.  The scalar power
spectrum has a step like feature with a burst of oscillations
connecting the two levels of the step (see, for instance, Fig.~3 of
Ref.~\cite{Martin:2011sn}). The flat regions of the step reflect the
two epochs of slow roll, while the oscillations in between arise as a 
result of the period of fast roll.

\par

Naively, one would have expected that the scalar bi-spectrum too would
exhibit a similar behavior, viz. that it would turn scale invariant
when slow roll has been restored. However, strikingly, when considered 
without adequate care, it is found that the bi-spectrum grows linearly 
with the wavenumbers at small scales (see,
Refs.~\cite{Arroja:2011yu,Arroja:2012ae}; in this context, also see
Figs.~7 and~11 in Ref.~\cite{Hazra:2012yn} and the recent work 
Ref.~\cite{Romano:2014kla}). From a theoretical point
of view, evidently, it is imperative to firmly establish the predictions
of the model and settle upon the correct behavior in a physically relevant 
and realistic situation. It is also worth noting here that, given the 
extent of accuracy of the measurements of non-Gaussianities by Planck, 
upon comparing with the data, the two behavior mentioned above would probably 
lead to very different constraints on the Starobinsky model. With these 
motivations in mind, in this work, as we have already pointed out, we 
intend to revisit the issue. 

\par

Clearly, the fact that the scale invariance of the bi-spectrum is not restored 
on large scales must be unphysical and, in this paper, we shall show that 
this arises due to the discontinuity in the second derivative of the potential. 
In fact, the point that the growing term is indeed unrealistic could have 
been easily guessed from the very beginning, since it exactly corresponds 
to a well-known and well-studied situation which was investigated long ago 
in the context of particle production by time-dependent, classical, 
gravitational fields~\cite{Ford:1986sy}. 
In what follows, we shall quickly recall the main results and conclusions
arrived at in the earlier work, as the phenomenon closely resembles the 
behavior of the bi-spectrum encountered in the Starobinsky model. 

\par

The earlier work~\cite{Ford:1986sy} considers a scalar field, say, $\psi$, 
that is non-minimally coupled to gravity, and is evolving in a spatially, 
flat, Friedmann-Lema\^{\i}tre-Robertson-Walker (for convenience, simply 
FLRW, hereafter) metric. Such a scalar field is governed by the following 
equation of motion: 
\begin{equation}
\l(\square-\xi\, R\r)\,\psi=0,
\end{equation}
where $R$ denotes the scalar curvature, while $\xi$ an arbitrary constant. 
In a time-dependent background such as the FLRW universe, it is common
knowledge that, upon quantization, pairs of particles associated with the 
scalar field~$\psi$ will, in general, be produced, provided the coupling is 
{\it not}\/ conformal, i.e. $\xi \neq 1/6$. Let $a(\eta)$ denote the scale 
factor of the FRLW universe, with $\eta$ being the conformal time 
coordinate. Upon Fourier transforming the scalar field and redefining the 
Fourier modes, say,  $\psi_{\bm k}$, as $\psi _{\bm k}\equiv \mu_{\bm k}/a(\eta)$, 
one finds that the differential equation satisfied by $\mu_{\bm k}$ can be
written as
\begin{equation}
\mu_{\bm k}''+k^2\,\mu_{\bm k}
={\mathcal V}_{\bm k}(\eta)\,\mu_{\bm k},
\end{equation}
where an overprime represents differentiation with respect to the conformal 
time $\eta$, while $k$ denotes the comoving wavenumber. 
The quantity ${\mathcal V}_{\bm k}(\eta)$ is given by 
\begin{equation}
{\mathcal V}_{\bm k}(\eta)\equiv 
\l(1-6\,\xi\r)\,\f{a''}{a}=\l(\f{1}{6}-\xi\r)\,a^2\,R.
\end{equation}
The above differential equation for $\mu_{\bm k}$ can also be cast as an 
integro-differential equation as follows:
\begin{equation}
\label{eq:eomintegro}
\mu_{\bm k}(\eta)=\f{{\rm e}^{-i\,k\,\eta}}{\sqrt{2\,k}}
+\frac{1}{k}\,\int _{-\infty}^{\eta}\d\tau\;{\mathcal V}_{\bm k}(\tau)\,
\sin\left[k\,\l(\eta-\tau\r)\right]\mu_{\bm k}(\tau). 
\end{equation}
At early stages of the expansion, the mode function can be expected to 
behave as $\mu_{\bm k} \rightarrow {\rm e}^{-i\,k\,\eta }/\sqrt{2\,k}$, 
which essentially corresponds to choosing the field to be in the vacuum
state initially. 
At late times, one has $\mu_{\bm k}\rightarrow ({\mathcal A}_{\bm k}/
\sqrt{2\,k})\, {\rm e}^{-ik\eta}+({\mathcal B}_{\bm k}/\sqrt{2\,k})\, 
{\rm e}^{i\,k\,\eta}$, where ${\mathcal A}_{\bm k}$ and 
${\mathcal B}_{\bm k}$ are the standard Bogoliubov coefficients that 
relate the modes at different times.
Then, using Eq.~(\ref{eq:eomintegro}), one can approximate the Bogoliubov
coefficient ${\mathcal B}_{\bm k}$ at very late times to be
\begin{equation}
\label{eq:betak}
{\mathcal B}_{\bm k}\simeq \frac{i}{2\,k}\,\int _{-\infty}^{\infty}\d\tau\;
{\mathcal V}_{\bm k}(\tau)\, {\rm e}^{-2\,i\,k\,\tau}.
\end{equation}
This expression, in turn, permits one to evaluate the energy 
density of the created particles, which is arrived at by 
calculating the integral~\cite{Ford:1986sy}
\begin{equation}
\rho=\f{1}{2\,\pi^2\,a^4}\,\int_0^{\infty}\d k\, k^3\;
\vert {\mathcal B}_{\bm k}\vert^2.
\end{equation}

\par

Let us now consider the case wherein there is a sharp transition from 
a phase of de Sitter inflation to a radiation dominated era. Let the 
transition take place at the conformal time, say,~$\eta_\ast$. In this 
scenario, the scalar curvature $R$ is non-zero, but constant 
(being related to the constant Hubble parameter during the de Sitter 
phase) for $\eta<\eta_\ast$, while $R$ vanishes for $\eta>\eta_\ast$ (i.e. 
during the radiation dominated epoch). 
The integral~(\ref{eq:betak}) can be carried out explicitly in such a case 
and, one obtains that, ${\mathcal B}_{\bm k}=2\,(1-6\,\xi)\;\Gamma(-1,2\,i
\,k\,\eta_\ast)$, where $\Gamma(b,z)$ is the incomplete Euler 
function~\cite{Gradshteyn:1965aa,Abramovitz:1970aa}. For large 
values of $k$, one finds that $\vert {\mathcal B}_{\bm k}\vert ^2 \propto 
k^{-4}$, with the result that the corresponding energy density $\rho$
diverges logarithmically. However, as discussed in the original 
work~\cite{Ford:1986sy}, this conclusion is unphysical, and it is just an 
artifact of the abruptness of the transition from the de Sitter phase to 
the epoch of radiation domination. Indeed, if we now `regularize' the 
transition, for instance, by smoothening out the quantity 
${\mathcal V}_{\bm k}(\eta)$ to be, say, 
${\mathcal V}_{\bm k}(\eta)=2\,(1-6\,\xi)/(\eta^2+\eta_\ast^2)$, then the 
coefficient ${\mathcal B}_{\bm k}$ is found to be 
\begin{equation}
{\mathcal B}_{\bm k}
=-\frac{i\,\pi}{k\,\eta_\ast}\,{\rm e}^{2\,k\,\eta_\ast}.
\end{equation}
In other words, one obtains an exponential cut-off in the spectrum, 
i.e. $\vert {\mathcal B}_\vk\vert^2$, of created particles (note that 
$\eta_\ast$ is negative), which occurs as a result of smoothening out 
the sharp transition. If we now calculate the corresponding energy 
density, then we arrive at a finite result, viz. 
$\rho=(1-6\,\xi)^2/(32\, a^4\,\eta_\ast^4)$. This unambiguously illustrates 
the point that the original logarithmic divergence was indeed an artifact 
and, upon modeling the transition more realistically, one obtains a  
result that is perfectly finite and physical.

\par

In the same manner, the indefinite growth of the bi-spectrum at small
scales in the Starobinsky model ought to be just an artifact and should
be considered to be unphysical. In this work, focusing on the equilateral 
limit, we shall analytically investigate the effects of smoothening out 
the discontinuity in the derivative of the potential on the scalar 
bi-spectrum. We shall also compare the analytical results with the numerical 
results from the code Bi-spectra and Non-Gaussianity Operator or, simply, 
BINGO, which we had recently put together to compute the scalar bi-spectrum
in inflationary models involving the canonical scalar field~\cite{Hazra:2012yn}. 
As we shall illustrate, in the case of the bi-spectrum, smoothening out the 
discontinuity restores the scale invariance of the bi-spectrum at suitably 
large wavenumbers, depending on the extent of the smoothening. This
allows us to conclude that the continued growth in the bi-spectrum at small 
scales, as was found earlier, can be attributed to the unrealistic assumption 
that the discontinuity in the derivative of the potential can be arbitrarily 
sharp.

\par

The remainder of this paper is organized as follows. In the following 
two sections, we shall highlight a few essential aspects of the 
Starobinsky model and discuss the dominant contribution to the scalar 
bi-spectrum (in the equilateral limit) which arises due to the discontinuity 
in the first derivative of the potential in the model. 
In Sec.~\ref{sec:esd}, we shall smoothen the discontinuity 
in a simple manner, which allows one to obtain the modes during the transition, 
and evaluate the corresponding contribution to the scalar bi-spectrum.
We shall see that even the simplest of smoothening curtails the growth of 
the bi-spectrum on small scales. 
In Sec.~\ref{sec:generalsmooth}, focusing on the limit of large wavenumbers,
we shall discuss the effects of a more generic smoothening of the potential.
We shall analytically illustrate that, if the potential is smoothened 
sufficiently, it ensures that the corresponding contributions to the 
bi-spectrum prove to be insignificant at suitably small scales.
In Sec.~\ref{sec:cnr}, we shall compare the analytical expressions we obtain 
with the numerical results from BINGO.
We shall conclude in Sec.~\ref{sec:d} with a few general remarks.

\par

Note that, we shall assume the background to be the spatially flat,
FLRW line-element, which is described by the scale factor~$a$ 
and the Hubble parameter $H$. Also, we shall work with units such 
that $c=\hbar=1$, and we shall set $\Mp^2 = (8\, \pi\, G)^{-1}$.  
Moreover, $t$ shall denote the cosmic time coordinate, and we shall 
represent differentiation with respect to~$t$ by an overdot.
As we have already mentioned, $\eta$ represents the conformal time 
coordinate, while an overprime denotes differentiation with respect 
to~$\eta$.
Further, $N$ shall denote the number of e-folds. 
Lastly, a plus sign, a zero or a minus sign in the sub-script or 
the super-script of any quantity shall denote its value or 
contribution before, during and after the field crosses the
discontinuity in the derivative of the potential, respectively.


\section{Essential aspects of the Starobinsky model}

The Starobinsky model involves a canonical scalar field and it
consists of a linear potential with a sudden change in its slope at a
given point~\cite{Starobinsky:1992ts}. The potential that describes
the model can be written as follows:
\begin{equation}
V(\phi) 
= \l\{\begin{array}{ll}
\displaystyle
V_0 + A_{+}\, \l(\phi-\phi_0\r)\ & {\rm for}\ \phi>\phi_0,\\
\displaystyle
V_0 + A_{-}\, \l(\phi-\phi_0\r)\ & {\rm for}\ \phi<\phi_0.
\end{array}\r.\label{eq:p-sm} 
\end{equation}
Evidently, while the value of the scalar field where the slope, i.e.
$V_\phi\equiv \d V/\d\phi$, changes abruptly is~$\phi_0$, the slope 
of the potential above and below $\phi_{0}$ are given by $A_{+}$ and 
$A_{-}$, respectively.  Moreover, the quantity $V_{0}$ denotes the 
value of the potential at $\phi=\phi_0$. In this section, we shall 
highlight a few important points relating to the evolution of the 
background, in particular, the behavior of the slow roll parameters, 
in the Starobinsky model. We shall also discuss the behavior of the 
modes describing the curvature perturbation before and after the 
field crosses the point~$\phi_0$.


\subsection{Evolution of the background}

An important assumption of the Starobinsky model is that the value of
$V_0$ is sufficiently large that it dominates the energy of the scalar
field as it rolls down the potential across~$\phi_0$. As a result, the
behavior of the scale factor proves to be essentially that of de
Sitter. This, in turn, implies that the first slow parameter,
viz. $\epsilon_1= -{\dot H}/H^2$, remains much smaller than unity
throughout the evolution, even as the field crosses the discontinuity
in the potential.  In fact, the first slow roll parameter before and
after the transition, i.e. when the field crosses~$\phi_0$, can be
shown to be~\cite{Martin:2011sn}
\begin{eqnarray}
\epsilon_{1+} &\simeq& \f{A_+^2}{18\, \Mp^2\,H_0^4},\\
\epsilon_{1-}
&\simeq&\f{A_-^2}{18\,\Mp^2\,H_0^4}\,
\left[1-\f{\Delta A}{A_-}\, {\rm e}^{-3\,\l(N-N_{0}\r)}\right]^2,
\end{eqnarray}
respectively, where $H_0$ is a constant that is determined by the
relation $H_0^2\simeq V_0/(3\, \Mpl^2)$, while $N_0$ denotes the
e-fold at the transition, and $\Delta A\equiv A_--A_+$.

\par

Before the transition, the second slow roll parameter, viz. $\epsilon_2=
\d\ln \epsilon_1/\d N$, is determined by the slow roll 
approximation and is found to be $\epsilon_{2+}\simeq 4\,\epsilon_{1+}$.
However, as the field crosses~$\phi_0$, the change in the slope causes a 
short period of deviation from slow roll.
After the transition, the second slow roll parameter $\epsilon_{2}$ is found 
to be~\cite{Martin:2011sn}
\begin{equation}
\epsilon_{2-}
\simeq \f{6\,\Delta A}{A_-}\, 
\frac{{\rm e}^{-3\,\l(N-N_{0}\r)}}
{1-\l(\Delta A/A_{-}\r)\, {\rm e}^{-3\,\l(N-N_{0}\r)}}
+4\, \epsilon_{1-}.
\end{equation} 
It is clear that $\epsilon_{2-}$ turns large immediately after the transition 
and, when slow roll is restored eventually, one finds that $\epsilon_{2-}\simeq
4\,\epsilon_{1-}$, just as one would expect.

\par

As we shall discuss in the following section, the dominant
contribution to the scalar bi-spectrum arises due to the so-called
fourth term in the Maldacena formalism (in this context, see, for
instance, Refs.~\cite{Martin:2011sn,Hazra:2012kq,Hazra:2012yn}).  This
contribution involves the time derivative of the second slow roll
parameter~$\epsilon_2$.  Upon using the background equations, one can
show that ${\dot \epsilon}_2$ can be written
as~\cite{Martin:2011sn,Arroja:2011yu,Arroja:2012ae}
\begin{equation}
{\dot \epsilon}_{2}
=-\f{2\, V_{\phi\phi}}{H}
+12\,H\,\epsilon_{1}-3\,H\,\epsilon_{2}-4\,H\,\epsilon_{1}^2
+5\,H\,\epsilon_{1}\,\epsilon_{2}-\frac{H}{2}\, \epsilon_{2}^2,
\label{eq:e2-ee}
\end{equation}
where $V_{\phi\phi}\equiv \d^2V/\d\phi^2$, and we should stress here
that this expression is an exact one. In the case of the Starobinsky
model, due to the discontinuity in the slope $V_\phi$ of the potential, 
clearly, the first term in the expression for ${\dot \epsilon}_2$ above, 
which involves the second derivative of the potential, will lead to a 
Dirac delta function. The contribution to ${\dot \epsilon}_{2}$ due 
to this specific term can then be written as
\begin{equation}
{\dot \epsilon}_{2}
\simeq\f{2\; \Delta A}{H_0}\; \delta^{(1)}(\phi-\phi_0)
=\f{6\; \Delta A}{A_+\, a_0}\; \delta^{(1)}(\eta-\eta_0).\label{eq:e2d}
\end{equation}
In fact, on large wavenumbers, as we shall soon discuss, it is this
particular term that was found to lead to the dominant contribution to
the scalar bi-spectrum~\cite{Arroja:2011yu,Arroja:2012ae}, if one works
in the limit where the discontinuity in $V_\phi$ is infinitely
sharp.


\subsection{Evolution of the perturbations}

Let us now turn to briefly discuss the behavior of the modes
describing the scalar perturbations in the Starobinsky model.

\par

Recall that, the Fourier modes of the curvature perturbations, say,
$f_{\bm k}(\eta)$, are governed by the differential
equation~\cite{Mukhanov:1990me}
\begin{equation}
f_{\bm k}''+2\, \f{z'}{z}\, f_{\bm k}' + k^{2}\, f_{\bm k}=0,\label{eq:defk}
\end{equation}
where $z =a\,\Mp\,\sqrt{2\,\epsilon_1}$. 
In terms of the Mukhanov-Sasaki variable, $v_{\bm k}=z\, f_{\bm k}$, the above 
equation for $f_{\bm k}$ reduces to
\begin{equation}
v_{\bm k}''+\l(k^2-\f{z''}{z}\r)\, v_{\bm k}=0.\label{eq:devk}
\end{equation}
The `effective potential' $z''/z$ that appears in this differential 
equation can be written in terms of the slow roll parameters as follows:
\begin{equation}
\frac{z''}{z}
={\cal H}^2\, \l(2-\epsilon_1+\f{3\, \epsilon_{2}}{2}
+\f{\epsilon_2^2}{4}
-\f{\epsilon_1\, \epsilon_{2}}{2}
+\f{\epsilon_2\, \epsilon_3}{2}\r),\label{eq:poteff}
\end{equation}
where ${\cal H}\equiv a'/a=a\, H$ is the conformal Hubble parameter,
while $\epsilon_3$ denotes the third slow roll parameter given by
\begin{equation}
\epsilon_3=\f{\d\,{\rm ln}\,\epsilon_2}{\d N}
=\f{{\dot \epsilon}_2}{H\, \epsilon_2}.\label{eq:e3}
\end{equation}
Also, it should be emphasized that the above expression for $z''/z$ is 
exact, and no approximation has been made in arriving at it.

\par

In the Starobinsky model, due to certain cancellations that occur
under the approximations of interest, one finds that the quantity
$z''/z$ reduces to $2\, {\cal H}^2$ before {\it as well as}\/ after the
transition~\cite{Starobinsky:1992ts,Martin:2011sn,Arroja:2011yu,Arroja:2012ae}.
This basically corresponds to the de Sitter limit, which then implies
that the Mukhanov-Sasaki variable $v_{\bm k}$ during these regimes is
essentially given by the conventional Bunch-Davies
solutions~\cite{Bunch:1978yq}. However, it should be clear from the
expressions~(\ref{eq:poteff}), (\ref{eq:e3}) and~(\ref{eq:e2d}) that,
at the transition, it is the last term involving the quantity
$\epsilon_3$ in $z''/z$ above which will dominate. One finds that the
corresponding effective potential is described by a Dirac delta
function {\it at}\/ the transition, and is given
by~\cite{Martin:2011sn}
\begin{eqnarray}
\frac{z''}{z}
&\simeq& \f{{\cal H}^2\,\epsilon_2\, \epsilon_3}{2}
=\f{{\cal H}^2\, {\dot \epsilon}_2}{2\,H}
=a_0^2\; \Delta A\;\delta^{(1)}\,\l(\phi -\phi_0\r)\nn\\
&=& \f{a_0^2\, \Delta A}{\vert \d \phi/\d \eta\vert_{\eta_0}}\;
\delta^{(1)}\,\l(\eta -\eta_0\r)
=\frac{3\,a_0\,H_{0}\,\Delta A}{A_{+}}\; 
\delta^{(1)}\l(\eta -\eta_0\r),\label{eq:effpot-at}
\end{eqnarray}
where $\eta_0$ and $a_0$ denote the conformal time and the scale
factor at the transition. We should clarify that, while the strictly
de Sitter term, viz. $z''/z \simeq 2\, {\cal H}^2$, remains, it is the
above term which a priori dominates {\it at}\/ the transition.

\par

Due to slow roll, before the transition, the modes $v_{\bm k}$ can be
described to a good approximation by following de Sitter solution:
\begin{equation}
v_\vk^{+}(\eta)
=\f{1}{\sqrt{2\, k}}\, 
\l(1-\frac{i}{k\,\eta}\r)\, {\rm e}^{-i\,k\,\eta}.\label{eq:vk-bt}
\end{equation}
Though slow roll is indeed restored at late times, due to the intervening 
epoch of fast roll, post-transition, the modes~$v_\vk$ do not remain in 
the Bunch-Davies vacuum.
Hence, after the transition, the solution to~$v_\vk$ takes the general 
form
\begin{equation}
v_\vk^{-}(\eta)
=\frac{\alpha_\vk}{\sqrt{2\, k}}\, 
\l(1-\frac{i}{k\,\eta}\r){\rm e}^{-i\,k\,\eta}
+\frac{\beta_\vk}{\sqrt{2\, k}}\, 
\l(1+\frac{i}{k\,\eta}\r){\rm e}^{i\,k\,\eta},
\label{eq:vk-at}
\end{equation}
where $\alpha_\vk$ and $\beta_\vk$ are the standard Bogoliubov
coefficients. The expression~(\ref{eq:effpot-at}) for $z''/z$ then
leads to the following matching conditions on the modes $v_\vk$ and
their derivatives $v_\vk'$ at the transition:
\begin{equation}
v_\vk^{-}\l(\eta_0\r) = v_\vk^{+}\l(\eta_0\r).
\end{equation}
and
\begin{equation}
v_\vk^{-}{}'\l(\eta_0\r) 
- v_\vk^{+}{}'\l(\eta_0\r)
=\f{3\,a_0\,H_0\,\Delta A}{A_{+}}\; v_\vk^+\l(\eta_0\r).
\end{equation}
These conditions then allow us to determine the Bogoliubov 
coefficients $\alpha_\vk$ and $\beta_\vk$, which can be obtained
to be
\begin{eqnarray}
\alpha_\vk 
&=& 1+\frac{3\,i\,\Delta A}{2\,A_{+}}\;\frac{k_0}{k}\,
\left(1+\frac{k_0^2}{k^2}\right),
\label{eq:alphak-sm}\\
\beta_\vk 
&=& -\frac{3\,i\,\Delta A}{2\,A_+}\;\f{k_0}{k}\,
\l(1+\frac{i\, k_0}{k}\r)^2\, {\rm e}^{2\,i\,k/k_{0}},
\label{eq:betak-sm}
\end{eqnarray}
where $k_0\equiv -1/\eta_{0}=a_0\,H_0$ corresponds to the mode that
leaves the Hubble radius at the transition.

\par

One can arrive at the corresponding expressions for the modes $f_\vk$ and
the derivative $f_\vk'$ before and after the transition from the above 
expressions for $v_\vk$ and its time derivative $v_\vk'$.
Before the transition, the mode $f_\vk$ and the derivative $f_\vk'$ are
given by
\begin{equation}
f_\vk^{+}(\eta)
=\frac{i\, H_0}{2\, \Mp\, \sqrt{{k^3}\,\epsilon_{1+}}}\,
\l(1+i\,k\,\eta\right)\,{\rm e}^{-i\,k\,\eta},\label{eq:fk-bt}
\end{equation}
and
\begin{equation}
f_\vk^{+}{}'(\eta)
=\frac{i\, H_0}{2\, \Mp\, \sqrt{{k^3}\,\epsilon_{1+}}}\,
\l[-{\cal H}\,\l(\epsilon_{1+}+\f{\epsilon_{2+}}{2}\r)\,
\l(1+i\,k\,\eta\r)
+k^2\,\eta\r] {\rm e}^{-i\,k\,\eta}.\label{eq:fkp-bt}
\end{equation}
Whereas, after the transition, one finds that 
\begin{eqnarray}
f_\vk^{-}(\eta)
=\frac{i\,H_0\,\alpha_\vk}{2\,\Mp\,\sqrt{{k^3}\,\epsilon_{1-}}}\,
\l(1+i\,k\,\eta\r)\, {\rm e}^{-i\,k\,\eta}
-\,\frac{i\,H_0\,\beta_\vk}{2\,\Mp\,\sqrt{{k^3}\,\epsilon_{1-}}}
\l(1-i\,k\,\eta\right)\, {\rm e}^{i\,k\,\eta}\label{eq:fk-at}
\end{eqnarray}
and
\begin{eqnarray}
f_\vk^{-}{}'(\eta)
&=&\frac{i\,H_0\,\alpha_\vk}{2\,\Mp\,\sqrt{{k^3}\epsilon_{1-}}}
\l[-{\cal H}\,\l(\epsilon_{1-}+\f{\epsilon_{2-}}{2}\r)\,\l(1+i\,k\,\eta\r)
+k^2\,\eta\r]
{\rm e}^{-i\,k\,\eta}\nn\\
&-&\frac{i\,H_0\,\beta_\vk}{2\,\Mp\,\sqrt{{k^3}\epsilon_{1-}}}
\l[-{\cal H}\,\l(\epsilon_{1-}+\f{\epsilon_{2-}}{2}\r)\,\l(1-i\,k\,\eta\right)
+k^2\,\eta\r]{\rm e}^{i\,k\,\eta}.\label{eq:fkp-at}\nn \\
\end{eqnarray}

\par

Note that, unlike the case of the Mukhanov-Sasaki
equation~(\ref{eq:devk}), the governing equation~(\ref{eq:defk}) for
$f_\vk$ involves only $z'/z$ rather than $z''/z$. It should also be
clear from the above arguments that $z'/z$ will involve the Heaviside
step function. This implies that the mode $f_\vk$ and its derivative
$f_\vk'$ are both continuous at the transition. As we shall discuss in
the next section, the most significant contribution to the dominant
term in the scalar bi-spectrum in the Starobinsky model shall depend
on the mode $f_\vk$ and the derivative $f_\vk'$ evaluated {\it at}\/
the transition. Because of their simpler structure, it proves to be
convenient to make use of the expressions~(\ref{eq:fk-bt}) and
(\ref{eq:fkp-bt}) for the mode $f_\vk$ and $f_\vk'$ before the
transition.  At the transition, these reduce to
\begin{equation}
f_\vk(\eta_0)
=\frac{i\, H_0}{2\, \Mp\, \sqrt{{k^3}\,\epsilon_{1+}}}\,
\l(1-\frac{i\,k}{k_0}\right)\,{\rm e}^{i\,k/k_0},
\end{equation}
and
\begin{eqnarray}
f_\vk'(\eta_0)
&=&-\frac{i\, H_0}{2\, \Mp\, \sqrt{{k^3}\,\epsilon_{1+}}}\,
\l[3\, \epsilon_{1+}\, k_0 \l(1-\frac{i\,k}{k_0}\r)
+\frac{k^2}{k_0}\r]\, {\rm e}^{i\,k/k_0}\nn\\
&\simeq& -\frac{i\, H_0}{2\, \Mp\, \sqrt{{k^3}\,\epsilon_{1+}}}\,
\frac{k^2}{k_0}\, {\rm e}^{i\,k/k_0},
\end{eqnarray}
where we have made use of the fact that $\epsilon_{2+}=4\, \epsilon_{1+}$ 
to obtain the first expression, and have ignored the term involving 
$\epsilon_{1+}$, as is done in the slow roll approximation, to arrive at
the second.


\section{The dominant contribution to the scalar bi-spectrum}
 
For simplicity, we shall focus on the equilateral limit in this work.
It is well known that, when deviations from slow roll occur, it is the
fourth term in the Maldacena formalism that leads to the dominant
contribution to the
bi-spectrum~\cite{Hazra:2012kq,Chen:2006xjb,Chen:2008wn,Hotchkiss:2009pj,Hannestad:2009yx,Flauger:2010ja,Adshead:2011bw,Adshead:2011jq}.
In the equilateral limit of our interest, the fourth term, which we
shall refer to as $G_4(k)$, is given by
\begin{equation}
G_4(k)= \Mp^2\, \l[f_\vk^3(\eta_{\rm e})\, \cG_4(k)
+f_\vk^{\ast}{}^3(\eta_{\rm e})\, \cG_4^{\ast}(k)\r],\label{eq:G4k}
\end{equation}
where $\ee$ denotes the end of inflation.
The quantity $\cG_4(k)$ is described by the integral
\begin{equation}
\cG_{4}(k)
=3\, i\int_{\ei}^{\ee} \d\eta\; a^3\,\epsilon_{1}\,
{\dot \epsilon}_{2}\, f_{\vk}^{\ast}{}^{2}\, f_{\vk}'^{\ast},
\label{eq:cG4}
\end{equation}
where $\ei$ denotes a very early time, say, when the initial conditions 
are imposed on the perturbations.

\par

\begin{figure}[!t]
\begin{center}
\resizebox{420pt}{280pt}{\includegraphics{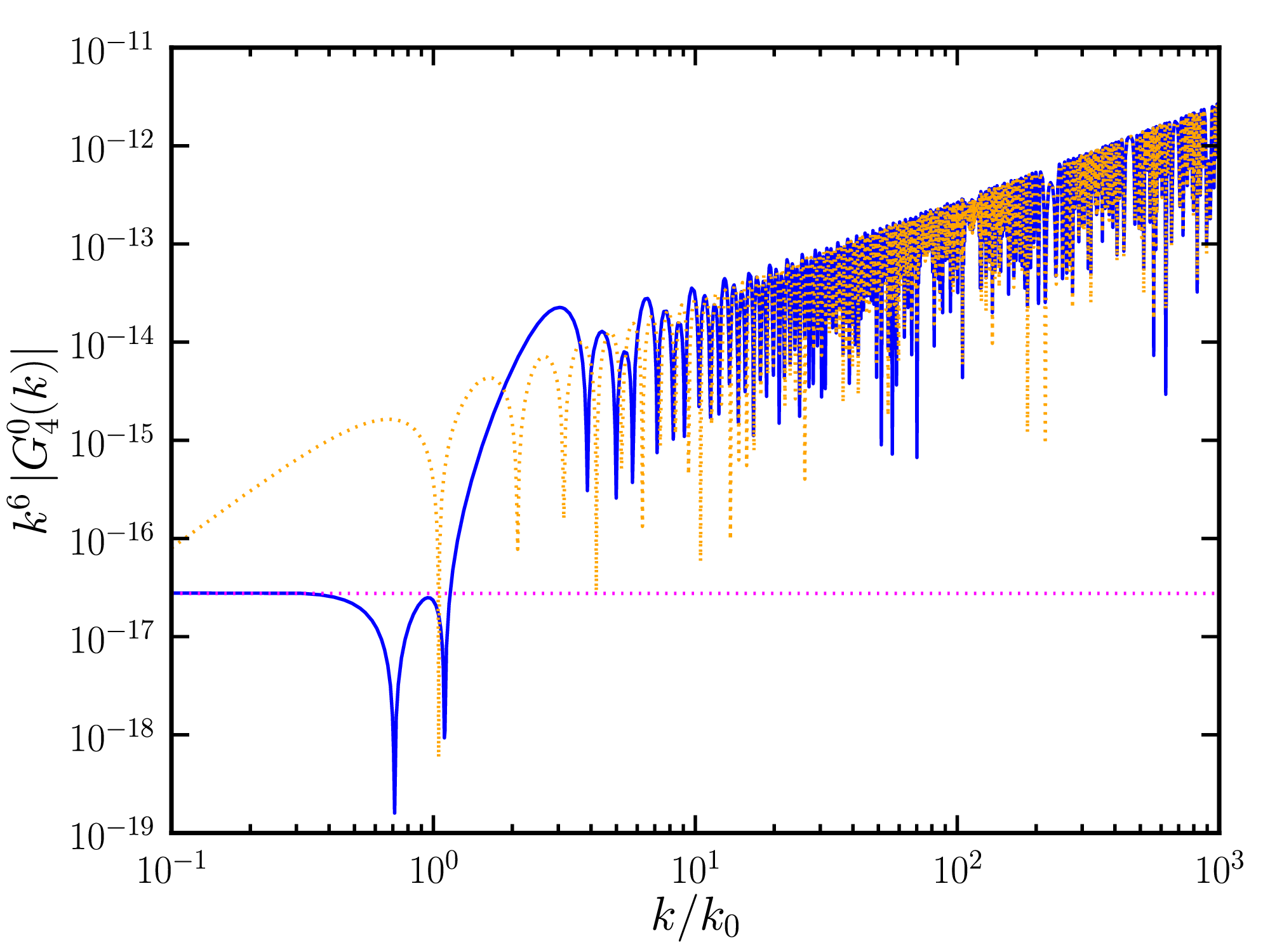}}
\end{center}
\vskip -10pt
\caption{\label{fig:oar}The behavior of the quantity $k^6\, \vert G_4^0(k)
\vert$ [cf. Eq.~(\ref{eq:G40-oar})] (in blue) as well as its behavior at small
(in magenta) and large (in orange) wavenumbers in the Starobinsky model.
The above plot corresponds to the following values of the parameters of
the Starobinsky model: $V_0=2.36\times 10^{-12}\, \Mpl^4$, $A_+=3.35\times 
10^{-14}\, \Mpl^3$, $A_-=7.26\times 10^{-15}\, \Mpl^3$ and $\phi_0=0.707\, 
\Mpl$. The linear growth at large wavenumbers is evident.}
\end{figure}

Recall that, in a generic situation, the complete expression for the 
quantity ${\dot \epsilon}_2$ is given by Eq.~(\ref{eq:e2-ee}).
As we have already discussed, in the Starobinsky model, the first term 
involving $V_{\phi\phi}$ in the exact expression for ${\dot \epsilon}_2$ 
leads to a delta function [cf. Eq.~(\ref{eq:e2d})]. It is then evident
from the integral~(\ref{eq:cG4}) that the corresponding contribution 
will be non-zero only {\it at}\/ the transition. 
Actually, the contributions due to all the other terms, i.e. apart 
from the term involving $V_{\phi\phi}$ in Eq.~(\ref{eq:e2-ee}), can be 
evaluated analytically (in this context, see Ref.~\cite{Martin:2011sn}).
However, we shall focus here only on the specific contribution due to the 
$V_{\phi\phi}$ term in $\dot{\epsilon}_2$, since it is this term that has 
been found to lead to the linear and indefinite growth on large wavenumbers
in the bi-spectrum~\cite{Arroja:2011yu,Arroja:2012ae}. We find that, with 
${\dot \epsilon}_2$ given by Eq.~(\ref{eq:e2d}), the quantity $\cG_{4}(k)$ 
can be written as
\begin{eqnarray}
\cG_{4}^0(k)
= \f{i\, \Delta A\, A_+\, k_0^2}{H_0^6\,\Mp^2}\, 
f_\vk^{\ast}{}^{2}(\eta_0)\, f_\vk'{}^\ast(\eta_0).
\end{eqnarray}
Towards the end of inflation, i.e. as $\eta\to 0$, the mode 
$f_\vk^-$ simplifies to
\begin{equation}
f_\vk^-(\eta_{\rm e})
=\frac{i\, H_0}{2\, \Mp\, \sqrt{{k^3}\,\epsilon_{1-}(\eta_{\rm e})}}\,
\l(\alpha_\vk-\beta _\vk\r),\label{eq:fk-lt}
\end{equation}
where $\epsilon_{1-}(\eta _{\rm e})$ denotes the value of the first
slow roll parameter at late times.
Upon using the above two expressions for $\cG_{4}^0(k)$ and 
$f_\vk^-(\eta_{\rm e})$ in the expression~(\ref{eq:G4k}) for $G_4(k)$, 
we find that we can write the contribution to the bi-spectrum due to the
transition as follows: 
\begin{eqnarray}
k^6\, G_{4}^{0}(k)
&=& -\frac{i\, \Delta A\, A_{+}}{64\, \Mp^6}\, \f{k_0}{k}\, 
\f{1}{\sqrt{\epsilon_{1+}^{3}\, \epsilon_{1-}^{3}\l(\eta_{\rm e}\r)}}\nn\\
& &\times\,\Biggl\{3\, i\,\biggl[\l(\alpha_\vk^2\, {\tilde \beta}_\vk
+\alpha_\vk\, {\tilde \beta}_\vk^2\r)\, {\cal C}(k)
+\l(\alpha_\vk^{\ast}{}^2\, {\tilde \beta}_\vk^{\ast}
+\,\alpha_\vk^{\ast}\, {\tilde \beta}_\vk^{\ast}{}^2\r)\, 
{\cal C}^{\ast}(k)\biggr]\,{\rm sin}\, \l(\f{k}{k_0}\r)\nn\\
& &\,- 3\,\biggl[\l(\alpha_\vk^2\, {\tilde \beta}_\vk
-\alpha_\vk\, {\tilde \beta}_\vk^2\r)\, {\cal C}(k)
- \l(\alpha_\vk^{\ast}{}^2\, {\tilde \beta}_\vk^{\ast}
-\alpha_\vk^{\ast}\, {\tilde \beta}_\vk^{\ast}{}^2\r)\, 
{\cal C}^{\ast}(k)\biggr]\,
{\rm cos}\, \l(\f{k}{k_0}\r)\nn\\
& &-\,i\, \l[\l(\alpha_\vk^3+{\tilde \beta}_\vk^3\r)\, {\cal C}(k)
+\l(\alpha_\vk^{\ast}{}^3+{\tilde \beta}_\vk^{\ast}{}^3\r)\, {\cal C}^{\ast}(k)\r]\,
{\rm sin} \l(\f{3\,k}{k_0}\r)\nn\\
& &+\, \l[\l(\alpha_\vk^3-{\tilde \beta}_\vk^3\r)\, {\cal C}(k)
-\l(\alpha_\vk^{\ast}{}^3-{\tilde \beta}_\vk^{\ast}{}^3\r)\, {\cal C}^{\ast}(k)\r]\,
{\rm cos} \l(\f{3\,k}{k_0}\r)\Biggr\},\label{eq:G40-oar}
\end{eqnarray}
where ${\tilde \beta}_\vk =\beta_\vk\, {\rm e}^{-2\,i\,k/k_0}$ and the quantity 
${\cal C}(k)$ is given by
\begin{equation}
{\cal C}(k)
=\l(1+\frac{i\,k}{k_0}\r)^2.
\end{equation}
As $k/k_0\to 0$, we find that $G_{4}^{0}(k)$ behaves as
\begin{equation}
\lim _{k/k_0\to 0}
k^6\, G_{4}^{0}(k)
=\f{-\,27\,\Delta A\, A_-^3\, H_0^6}{8\, A_+^5\,\Mp^3\,
\sqrt{2\,\epsilon_{1-}^{3}\l(\eta_{\rm e}\r)}},\label{eq:G40-oar-sk}
\end{equation}
while, in the limit $k/k_0\to \infty$, one obtains that
\begin{eqnarray}
\lim _{k/k_0\to \infty}
k^6\, G_{4}^{0}(k)
=\f{27\, \Delta A\, H_0^6}{8\, A_+^2\, \Mp^3\,
\sqrt{2\, \epsilon_{1-}^{3}\l(\eta_{\rm e}\r)}}\; \f{k}{k_0}\;
{\rm sin}\, \l(\f{3\,k}{k_0}\r).\label{eq:G40-oar-lk}
\end{eqnarray}
In Fig.~\ref{fig:oar}, we have plotted the absolute values of the exact 
result~(\ref{eq:G40-oar}) for the quantity $k^6$ times $G_{4}^{0}(k)$ 
as well as its asymptotic forms~(\ref{eq:G40-oar-sk}) 
and~(\ref{eq:G40-oar-lk}). 
Note the linear growth with $k$ at large
wavenumbers~\cite{Hazra:2012yn,Arroja:2011yu,Arroja:2012ae}. As we had
discussed earlier, one physically expects the bi-spectrum to turn
scale invariant for small scale modes that leave the Hubble radius at
late times, when slow roll has been reestablished.  However, one finds
here that the bi-spectrum continues to grow indefinitely with the
wavenumber.  Evidently, this can be attributed to the fact that the
potential contains an infinitely sharp transition, which can be 
considered to be unphysical, as discussed in the introductory section.
As we shall illustrate in the following sections, the indefinite growth 
disappears as one smoothens the transition.


\section{Effects of smoothening the discontinuity:~A simple
analytical treatment}\label{sec:esd}

In this section and the next, we shall analytically consider the effects 
of smoothening the discontinuity on the scalar bi-spectrum.  We shall
focus on the contribution due to the fourth term and we shall restrict
ourselves to the specific term in ${\dot \epsilon}_2$ (viz. the one
involving $V_{\phi\phi}$) that leads to the indefinite growth in the
scalar bi-spectrum.
 
\par

We shall first study the effects on the scalar bi-spectrum by smoothening
the discontinuity in a specific fashion that permits a relatively complete 
analytical treatment of the problem. Essentially, we shall replace the delta
function by one of its conventional representations.  Let us write
the delta function involved, viz. $\delta^{(1)}(\eta-\eta_0)$, in the
following fashion:
\begin{equation}
\delta^{(1)}(\eta-\eta_0)= \l\{\begin{array}{ll}
\displaystyle
0\ & {\rm for}\ \eta<\eta_-\\
\displaystyle
\f{1}{\varepsilon} & {\rm for}\ 
\eta_-<\eta<\eta_+,\\
\displaystyle
0\ & {\rm for}\ \eta>\eta_+,
\end{array}\r.\label{eq:rdd-fn} 
\end{equation}
where, for convenience, we have set 
\begin{equation}
\label{eq:defetapm}
\eta_\pm = \eta_0\pm \frac{\varepsilon}{2},
\end{equation}
with $\varepsilon$ being a small quantity (not to be confused with the
slow roll parameters) that determines the width and the height of
the transition.  Obviously, the limit $\varepsilon \to 0$ corresponds
to the original sharp transition.  In other words, instead of a
function of infinite height and infinitesimal width, we shall alter
the width and height suitably such that the area under the function is
unity, as is required.  In such a situation, in contrast to the
infinitely sharp transition wherein there had existed just two
domains, viz. the ones before and after the transition, there now
exists a third domain corresponding to the period of the transition.
It is then clear that, in the two original domains, i.e. when
$\eta<\eta_-$ and $\eta>\eta_+$, we have
$z''/z \simeq 2\, {\cal H}^2$, just as we had before.  Hence, the
earlier solutions for $v_\vk$, viz. (\ref{eq:vk-bt}) and
(\ref{eq:vk-at}) continue to remain valid during these domains.
However, during the transition, i.e. when $\eta_- < \eta< \eta_+$, we 
have
\begin{equation}
\f{z''}{z}\simeq  2\, {\cal H}^2 +\f{3\, a_0\, 
H_0\Delta A}{A_+\, \varepsilon }.
\end{equation}

\par

In order to be able to solve for the modes analytically corresponding
to the $z''/z$ above and also to be able to evaluate the integral
describing the quantity $\cG_4$ [see Eq.~(\ref{eq:cG4})], we shall
assume a few further points. Recall that, the delta function
encountered in $z''/z$ arises essentially due to its dependence on
${\dot \epsilon}_2$ [cf. Eq.~(\ref{eq:effpot-at})].  Therefore, by
altering the delta function, we have essentially modified the behavior
of $\epsilon_2'$ during the transition to be
\begin{equation}
\epsilon_2^{0}{}'=\f{6\, \Delta A}{A_{+}\,\varepsilon}.\label{eq:e2p-dt} 
\end{equation}
In such a case, clearly, during the transition, we would have 
\begin{equation}
\epsilon_2^0(\eta)
=\gamma\, (\eta-\eta_-) + 4\, \epsilon_{1+},\label{eq:e2-dt} 
\end{equation}
where, for convenience, we have set $\gamma=6\, \Delta A/(A_{+}\,\varepsilon)$.
Actually, such a modification would also result in a change in the behavior 
of the first slow roll parameter $\epsilon_1$ and, needless to add, the 
scale factor as well.  But, we shall assume that the scale factor continues 
to behave as that of de Sitter, and that the first slow roll parameter remains 
small and largely constant during the transition. As we shall see, these
assumptions allow us to arrive at a complete analytical form for the 
resulting bi-spectrum, with the expected limit as $\varepsilon \to 0$.

\par

Under the above assumptions, during the transition, the quantity $z''/z$ 
is given by 
\begin{equation}
\f{z''}{z}\simeq  \f{2}{\eta^2} +\f{3\, a_0\, 
H_0\, \Delta A}{A_+\, \varepsilon},
\end{equation}
and the corresponding solution to the Mukhanov-Sasaki equation can be
written as
\begin{equation}
v_\vk^{0}(\eta)
=\f{{\bar \alpha}_\vk}{\sqrt{2\, q}}\, 
\l(1-\frac{i}{q\,\eta}\r)\,{\rm e}^{-i\,q\,\eta}
+\frac{{\bar \beta}_\vk}{\sqrt{2\, q}}\, 
\l(1+\frac{i}{q\,\eta}\r)\,{\rm e}^{i\,q\,\eta},
\label{eq:vk-dt}
\end{equation}
where  ${\bar \alpha}_\vk$ and ${\bar \beta}_\vk$ denote the Bogoliubov 
coefficients during the transition, while
\begin{equation}
q^2= k^2-\f{3\,a_0\, H_0\, \Delta A}{A_+\, \varepsilon}.
\end{equation}
It is important to stress that, since $\Delta A<0$ for the parameter values 
of our interest (in this context, see the caption of Fig.~\ref{fig:oar}) 
and, as $a_0$ and $H_0$ are positive quantities, $q^2$ is a positive definite 
quantity.
The corresponding mode $f_\vk^{0}$ and its derivative $f_\vk^{0}{}'$ are 
given by
\begin{eqnarray}
f_\vk^{0}(\eta)
&=&\frac{i\,H_0\,{\bar \alpha}_\vk}{2\,\Mp\,\sqrt{{q^3}\,\epsilon_{1}^0}}\,
\l(1+i\,q\,\eta\r)\, {\rm e}^{-i\,q\,\eta}
-\frac{i\,H_0\,{\bar \beta}_k}{2\,\Mp\,\sqrt{{q^3}\,\epsilon_{1}^0}}\,
\l(1-i\,q\,\eta\r)\, {\rm e}^{i\,q\,\eta},\label{eq:fk-dt}
\end{eqnarray}
and
\begin{eqnarray}
f_\vk^{0}{}'(\eta)
&=&\frac{i\,H_0\,{\bar \alpha}_\vk}{2\,\Mp\,\sqrt{{q^3}\,\epsilon_{1}^0}}\,
\l[-{\cal H}\,\l(\epsilon_{1}^0+\f{\epsilon_{2}^0}{2}\r)\,
\l(1+i\,q\,\eta\r)+q^2\,\eta\r]\,{\rm e}^{-i\,q\,\eta}\nn\\
&-&\frac{i\,H_0\,{\bar \beta}_\vk}{2\,\Mp\,\sqrt{{q^3}\,\epsilon_{1}^0}}\,
\l[-{\cal H}\, \l(\epsilon_{1}^0+\f{\epsilon_{2}^0}{2}\r)\,
\l(1-i\,q\,\eta\r)+q^2\,\eta\r]\,{\rm e}^{i\,q\,\eta},\label{eq:fkp-dt}
\end{eqnarray}
where $\epsilon_{1}^0$ and $\epsilon_2^0$ represent the first two
slow roll parameters during the transition.

\par

The expressions for the Bogoliubov coefficients during the transition,
viz.  ${\bar \alpha}_\vk$ and ${\bar \beta}_\vk$, are obtained by
matching the modes $v_\vk$ and their derivatives $v_\vk'$ on either side 
at $\eta_-$. It should also be clear that the Bogoliubov coefficients
after the transition, i.e. $\alpha_\vk$ and $\beta_\vk$, will no more be
given by the original expressions [viz. Eqs.~(\ref{eq:alphak-sm}) and
(\ref{eq:betak-sm})], but will be modified.  They are arrived at by
matching the modes at $\eta_+$.  We find that the Bogoliubov
coefficients ${\bar \alpha}_\vk$ and ${\bar \beta}_\vk$ are given by
\begin{eqnarray}
{\bar \alpha}_\vk
&=&\f{1}{2\, \eta_-}\, \f{1}{(k\, q)^{3/2}}\; (k+q)\; 
(k\, q\, \eta_-+i\, k-iq)\;
{\rm e}^{-i\, (k-q)\, \eta_-},\label{eq:alpha1}\\
{\bar \beta}_\vk
&=&-\f{1}{2\, \eta_-}\, \f{1}{(k\, q)^{3/2}}\; (k-q)\; 
(k\, q\, \eta_--i\, k-iq)\;
{\rm e}^{-i\, (k+q)\, \eta_-}.\label{eq:beta1}
\end{eqnarray}
The Bogoliubov coefficients, say, $\alpha_\vk$ and 
$\beta_\vk$ in the domain
$\eta>\eta_+$ can be calculated to be
\begin{eqnarray}
\alpha_\vk
&=&\f{1}{2\, \eta_+}\, \f{1}{(k\, q)^{3/2}}\; 
\biggl[(k+q)\; (k\, q\, \eta_+-i\, k+i\,q)\;{\bar \alpha}_\vk\;
{\rm e}^{i\, (k-q)\, \eta_+}\nn\\
& &+\,(k-q)\; (k\, q\, \eta_+ + i\, k+i\,q)\;{\bar \beta}_\vk\;
{\rm e}^{i\, (k+q)\, \eta_+}\biggr],\label{eq:alpha-m}\\
\beta_\vk
&=&\f{1}{2\, \eta_+}\, \f{1}{(k\, q)^{3/2}}\; 
\biggl[(k-q)\; (k\, q\, \eta_+-i\, k-i\,q)\; {\bar \alpha}_\vk\;
{\rm e}^{-i\, (k+q)\, \eta_+}\nn\\
& &+\,(k+q)\; (k\, q\, \eta_+ + i\, k - i\,q)\; {\bar \beta}_\vk\;
{\rm e}^{-i\, (k-q)\, \eta_+}\biggr].\label{eq:beta-m}
\end{eqnarray}
One can easily show that, as $\varepsilon\to 0$, these expressions
simplify to the original expressions, viz. (\ref{eq:alphak-sm}) and
(\ref{eq:betak-sm}), for $\alpha_\vk$ and $\beta_\vk$.

\par

Recall that, our aim is to evaluate contribution to the bi-spectrum 
{\it during}\/ the transition, when it has been smoothened.
It is now a matter of substituting the mode~(\ref{eq:fk-dt}) and the 
corresponding derivative~(\ref{eq:fkp-dt}) in the expression~(\ref{eq:cG4}) 
and evaluating the integral involved from $\eta_-$ to $\eta_+$.
We find that, we can write ${\cal G}_4^0(k)$ as
\begin{equation}
{\cal G}_4^0(k)= 3\, i\; 
\l[{\bar \alpha}_\vk^{\ast}{}^3\, I_4^0(k) 
+{\bar \beta}_\vk^{\ast}{}^3\, I_4^0{}^{\ast}(k) 
+{\bar \alpha}_\vk^{\ast}{}^2\, {\bar \beta}_\vk^{\ast}\, J_4^0(k) 
+{\bar \alpha}_\vk^{\ast}{}\, 
{\bar \beta}_\vk^{\ast}{}^2\, J_4^0{}^{\ast}(k)\r],\label{eq:cG-dt}
\end{equation}
where the quantities $I_4^0(k)$ and $J_4^0(k)$ are described by the integrals
\begin{eqnarray}
I_4^0(k)&=&\f{i\, H_0\,\gamma}{16\, \Mpl^3\, q^3\, \sqrt{q^{3}\,\epsilon_1^0}}\,
\int_{\eta_-}^{\eta_+}\f{\d\eta}{\eta^3}\, {\rm e}^{3\, i\, q\, \eta}\,
\l(1-i\,q\,\eta\r)^2
\nn\\
& &\times\;
\biggl\{\l[\gamma\, \l(\eta-\eta_-\r)+4\,\epsilon_{1+}\r]\,
 \l(1-i\,q\,\eta\r)+2\,q^2\,\eta^2\biggr\},\\
J_4^0(k)&=&\f{-\,i\, H_0\,\gamma}{16\, \Mpl^3\, q^3\, \sqrt{q^{3}\,\epsilon_1^0}}\, 
\int_{\eta_-}^{\eta_+}\f{\d\eta}{\eta^3}\, {\rm e}^{ i\, q\, \eta}\,
\l(1-i\,q\,\eta\r)\nn\\
& &\times\biggl\{3\, \l[\gamma\, 
\l(\eta-\eta_-\r)+4\,\epsilon_{1+}\r]\, \l(1+q^2\,\eta^2\r)
+2\,q^2\,\eta^2\, \l(1-i\,q\,\eta\r)\nn\\
& &+\;4\, q^2\,\eta^2\, \l(1+i\, q\, \eta\r)\biggr\}.
\end{eqnarray}
We should mention that, in arriving at these integrals, we have ignored 
the term involving $\epsilon_1^0$ in the expression for 
$f_\vk^0{}'$ [the one within the square brackets in Eq.~(\ref{eq:fkp-dt})],
and we have made use of the expressions~(\ref{eq:e2p-dt}) and~(\ref{eq:e2-dt})
for $\epsilon_2^0{}'$ and $\epsilon_2^0$, respectively.
If we also ignore the term involving $\epsilon_2^0$ [within the square
brackets in Eq.~(\ref{eq:fkp-dt})], we find that the above integrals 
can be trivially integrated to arrive at the following 
results\footnote{We should clarify a point here. We find that the 
final results and conclusions we have presented below remain largely 
unaffected, even if we retain the term involving $\epsilon_2^0$.}:
\begin{eqnarray}
I_4^0(k)
&=&\f{i\, H_0\, \gamma}{16\, \Mpl^3\, q^3\, \sqrt{q^{3}\,\epsilon_1^0}}\;
{\rm e}^{3\,i\,q\,\eta_0}\;
\biggl\{-\l(\f{4\,q^3\,\eta_0}{3}+\f{28\,i\,q^2}{9}\r)\,
{\rm sin}\,\left(\frac{3\,q\,\varepsilon}{2}\right)\nn\\
& &+\,\f{2\,i\,q^3\,\varepsilon}{3}\, 
{\rm cos}\,\left(\frac{3\,q\,\varepsilon}{2}\right)\nn\\
& &+\, 2\, q^2\, {\rm e}^{-3\,i\,q\,\eta_0}\, \l[{\rm Ei}\,(3\,i\,q\,\eta_+)
-{\rm Ei}\,(3\,i\,q\,\eta_-)\r]\biggr\},\\
J_4^0(k)
&=&\f{-\,i\, H_0\, \gamma}{16\, \Mpl^3\, q^3\, \sqrt{q^{3}\,\epsilon_1^0}}\;
{\rm e}^{i\,q\,\eta_0}\,
\biggl\{-\l(4\,i\,q^2-4\,q^3\,\eta_0\r)\, 
{\rm sin}\,\left(\frac{q\,\varepsilon}{2}\right)\nn\\
& &-\,2\,i\,q^3\,\varepsilon\, 
{\rm cos}\,\left(\frac{q\,\varepsilon}{2}\right)
+ 6\, q^2\, {\rm e}^{-i\,q\,\eta_0}\, \l[{\rm Ei}\,(i\,q\,\eta_+)
-{\rm Ei}\,(i\,q\,\eta_-)\r]\biggr\},
\end{eqnarray}
where ${\rm Ei}(x)$ denotes the exponential integral 
function~\cite{Gradshteyn:1965aa,Abramovitz:1970aa}.

\par

The resulting contribution to the bi-spectrum can be obtained by
substituting the above results for $I_4^0(k)$ and $J_4^0(k)$ in the
expression~(\ref{eq:cG-dt}) for $\cG_4^{0}(k)$ and, in turn,
substituting the resulting $\cG_4^{0}(k)$ in the
expression~(\ref{eq:G4k}) and making use of the
behavior~(\ref{eq:fk-lt}) of the modes $f_\vk^-$ at late times [with
  $\alpha_\vk$ and $\beta_\vk$ now being given by
  Eqs.~(\ref{eq:alpha-m}) and (\ref{eq:beta-m})]. The complete
expression for $G_4^{0}(k)$ is quite long and unwieldy and, hence, we
will not write it down here.  However, its form in the limit of
$k/k_0\to\infty$, which is the behavior of our principal focus, can be
arrived at easily. One obtains that
\begin{eqnarray}
\lim _{k/k_0\to \infty}
k^6\, G_{4}^{0}(k)
&=&\f{27\, \Delta A\, H_0^6}{8\, A_+^2\, \Mp^3\,
\sqrt{2\,\epsilon_{1-}^{3}\l(\eta_{\rm e}\r)}}
\l[\f{2}{3\,\varepsilon\,k}\; 
{\rm sin}\, \l(\f{3\,k\,\varepsilon}{2}\r)\r]\,
\f{k}{k_0}\; {\rm sin}\, \l(\f{3\,k}{k_0}\r).\label{eq:G40-mar-lk}
\end{eqnarray}
It ought to be highlighted that, in the large $k$ limit, the only 
additional factor [as compared to the original result (\ref{eq:G40-oar-lk})] 
that arises due to the smoothening of the transition is the one that appears 
within the square brackets in the above expression.
Note that, the quantity $\varepsilon$ has dimensions of time, and the width
as well as the sharpness of the step are determined by the ratio $\varepsilon
/\vert\eta_0\vert$ or, equivalently, $\varepsilon\, k_0$.
If we write $\varepsilon=\kappa/k_0$, where $\kappa$ is a dimensionless
quantity, then we arrive at 
\begin{eqnarray}
\lim _{k/k_0\to \infty}
k^6\, G_{4}^{0}(k)
&=&\f{27\, \Delta A\, H_0^6}{8\, A_+^2\, \Mp^3\,
\sqrt{2\,\epsilon_{1-}^{3}\l(\eta_{\rm e}\r)}}
\l[\f{2\, k_0}{3\,\kappa\,k}\;
{\rm sin}\, \l(\f{3\,\kappa\,k}{2\, k_0}\r)\r]\,
\f{k}{k_0}\; {\rm sin}\, \l(\f{3\,k}{k_0}\r),\label{eq:G40-mar-lk-fv}
\end{eqnarray}
an expression that can be said to be the first important result of 
this paper.

\par

\begin{figure}[!t]
\begin{center}
\resizebox{420pt}{280pt}{\includegraphics{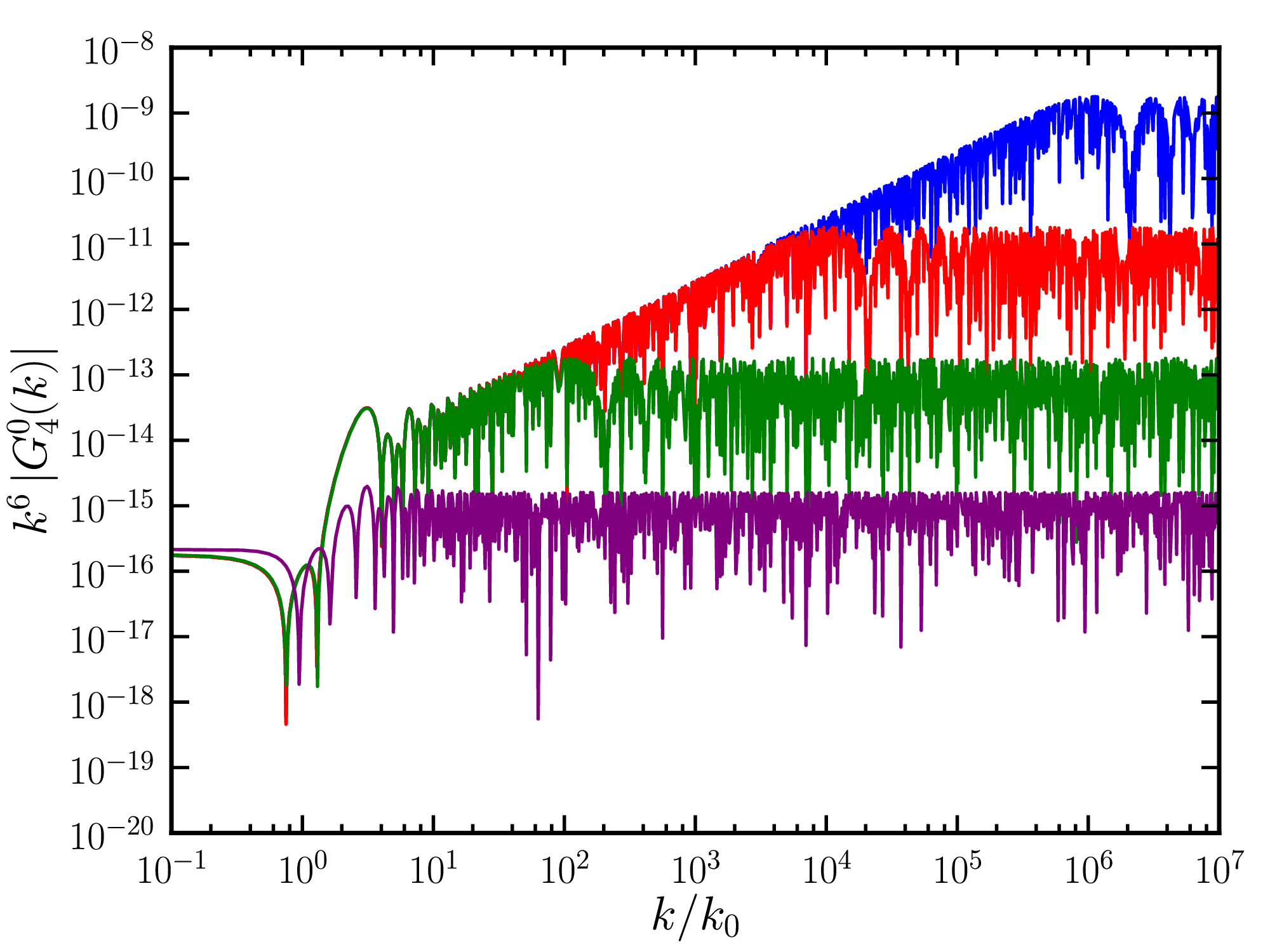}}
\end{center}
\vskip -10pt
\caption{\label{fig:mar}The behavior of the quantity $k^6\, \vert
  G_4^0(k)\vert$ in the Starobinsky model, when the delta function is
  represented by Eq.~(\ref{eq:rdd-fn}).  We have worked with
  the same values for the set of parameters that describe the original
  Starobinsky model (viz. $V_0$, $A_+$ and $A_-$) as in the last
  figure.  The different plots correspond to the following values of
  the dimensionless parameter $\kappa=\varepsilon\, k_0$: $10^{-6}$
  (in blue), $10^{-4}$ (in red), $10^{-2}$ (in green) and $1$ (in
  purple). It is evident that the smoothened step curtails the growth
  at $k/k_0\simeq \kappa^{-1}$.  Also, it is clear that the
  ratio of the scale invariant amplitudes at large $k/k_0$ is
  inversely proportional to the values of $\kappa$, thereby confirming
  the limiting behavior~(\ref{eq:G40-mar-lk-fv}) that we had arrived
  at analytically.}
\end{figure}

Three points need to be emphasized regarding the result we have
arrived at above. Firstly, it should be evident that the additional
factor [when compared to the original
expression~(\ref{eq:G40-oar-lk})] reduces to unity in the limit
$\varepsilon$ tends to zero, exactly as is required. This suggests
that the assumptions and methods we have adopted to smoothen the step
seem reasonable. Secondly, the above expression does not grow
indefinitely, but turns finite at large~$k$. It saturates at a scale
invariant amplitude that is inversely proportional to the value of
$\kappa$ and, moreover, the quantity turns scale invariant at
$k/k_0\simeq \kappa^{-1}$. In Fig.~\ref{fig:mar}, we have
plotted the complete expression for the absolute value of the quantity
$k^6\, G_{4}^{0}(k)$ for different values of~$\kappa$. A close look
clearly indicates that the figure completely corroborates the two
points we have made above. Thirdly and lastly, it is interesting to
compare the above result with the contribution to the bi-spectrum that
arises when all the other terms in ${\dot \epsilon_2}$, i.e. apart
from the term involving $V_{\phi\phi}$, are taken into account.  In
such a situation, one obtains that (in this context, see Eq.~(106) of
Ref.~\cite{Martin:2011sn})
\begin{eqnarray}
\lim _{k/k_0\to \infty}
k^6\, {\bar G}_{4}(k)
&=&\f{27\, \Delta A\, A_-\, H_0^6}{8\, A_+^3\, \Mp^3\,
\sqrt{2\,\epsilon_{1-}^{3}\l(\eta_{\rm e}\r)}}\, 
\cos\, \l(\f{3\,k}{k_0}\r).
\label{eq:G4ms}
\end{eqnarray}
If one neglects the trigonometric functions that are of order unity in
the above two expressions, one finds that
\begin{equation}
\lim _{k/k_0\to \infty} 
\frac{ k^6\, {\bar G}_{4}(k)}{k^6\, G_{4}^{0}(k)}
\simeq \frac{3\,\kappa}{2}\, \f{A_-}{A_+}
\simeq \frac{\Delta k}{k_0}\,\frac{A_-}{A_+},
\end{equation}
where we have set $\Delta k\simeq \varepsilon\, k_0^2 = \kappa\, k_0$, 
at first order in $\varepsilon$. This suggests that the contribution 
originating from the unphysical, growing term can be negligible provided 
$\Delta k/k_0\gg A_+/A_-$. In other words, if the transition is 
sufficiently smooth, then the growing term cannot rise to be too large. 
For the values of the parameters we have worked with in Figs.~\ref{fig:oar}
and~\ref{fig:mar}, viz. $A_+=3.35\times 10^{-14}\Mp^3$ and $A_-=7.26\times 
10^{-15}\Mp^3$, the condition we have arrived at above leads to $\Delta k
/k_0\simeq \kappa \gg 4.6$. It remains to 
be seen if this represents a reasonable value in the context of a realistic 
model. But, in any case, we have explicitly shown here that the bi-spectrum 
turns scale invariant on suitably small scales, when the step is smoothened.

\section{Smoothening the transition:~A more general treatment}
\label{sec:generalsmooth}

The calculation of the last section was based on a specific  
representation of the regularized Dirac delta function, as 
given by Eq.~(\ref{eq:rdd-fn}).
One may wonder as to what happens if we alter the fashion in 
which the transition is smoothened. 
In particular, it will be interesting to examine whether the plateau 
at small scales, as seen in Fig.~\ref{fig:mar}, generically appears 
whenever the transition is no longer infinitely sharp. 
In the previous section, we had chosen to work with the simple 
representation~(\ref{eq:rdd-fn}) of the regularized delta function, 
since it had permitted us to analytically determine the modes during 
the transition.
But, the modes prove to be difficult to obtain for a generic 
representation of the delta function.
However, as we are only interested in the small scale limit of the 
bi-spectrum, we find that, fortunately, it turns out to be possible 
to arrive at its behavior analytically in this regime in a simple 
manner, as we shall now describe.

\par

To begin with, note that, for very large wavenumbers, i.e. in the 
extreme small scale limit, the modes are completely unaffected
by the background.
As a result, in this limit, the modes $f_k^{0}$ and their time 
derivative $f_k^{0}{}'$ during the transition will be given by
\begin{eqnarray}
f_\vk^{0}(\eta)
&=&\frac{-H_0\,k\,\eta}{2\,\Mp\,\sqrt{{k^3}\,\epsilon_{1}^0}}\;
{\rm e}^{-i\,k\,\eta}\label{eq:fk-dt-ki}
\end{eqnarray}
and
\begin{eqnarray}
f_\vk^{0}{}'(\eta)
&=&\frac{i\,H_0\,k^2\,\eta}{2\,\Mp\,\sqrt{{k^3}\,\epsilon_{1}^0}}\,
{\rm e}^{-i\,k\,\eta},\label{eq:fkp-dt-ki}
\end{eqnarray} 
which are basically the behavior of the modes {\it before}\/ the 
transition.
Actually, for the extremely small scale modes, the above form will 
be valid {\it even after}\/ the transition.
In other words, the form of the modes remain unchanged throughout 
the evolution.
Moreover, it should be clear that this behavior is independent of 
the details of the transition.
Due to this reason, the Bogoliubov transformations turn trivial, 
with ${\bar \alpha}_\vk$ and $\alpha_\vk$ reducing to unity, while 
${\bar \beta}_\vk$ and $\beta_\vk$ vanish.
It is straightforward to check that the above-mentioned behavior are indeed
satisfied by the modes and the Bogoliubov coefficients for the specific form
of the regularized delta function representation considered in the previous
section.
For instance, as $k\to \infty$, one finds that ${\bar \alpha}_\vk$ and 
$\alpha_\vk$ [as given by Eqs.~(\ref{eq:alpha1}) and~(\ref{eq:alpha-m})] 
reduce to unity, whereas ${\bar \beta}_\vk$ and $\beta_\vk$ [as given by
Eqs.~(\ref{eq:beta1}) and~(\ref{eq:beta-m})] vanish, just as expected.
Further, since $q\to k$ for large $k$, the mode~(\ref{eq:fk-dt}) 
and its derivative~(\ref{eq:fkp-dt}) indeed reduce to the above forms 
for $f_k^{0}$ and $f_k^{0}{}'$. 

\par

Therefore, for the extreme small scale modes, we find that the quantity 
$\cG_4^{0}(k)$, i.e. the integral characterizing the contribution to the 
bi-spectrum during the transition, reduces to the following simple form:
\begin{equation}
{\cal G}_4^0(k)= \f{3\, H_0\, k}{8\, \Mp^3\, \sqrt{{k^3}\,\epsilon_{1}^0}}\,
\int_{\eta_-}^{\eta_+}\, \d \eta\; \eta\; \epsilon_2^{0}{}'\;
{\rm e}^{3\,i\,k\,\eta},\label{eq:cG40-lk}
\end{equation}
which is essentially the main result of this article. 
It is important to appreciate the point that this expression for 
applies to {\it any}\/ smooth transition in the small scale limit.
Note that, ${\cal G}_4^0(k)$ is basically the Fourier transform of 
the combination $\eta\; \epsilon_2^{0}{}'$.
Moreover, since $\alpha_\vk$ is unity, while $\beta_\vk$ vanishes
for small scales, the mode $f_\vk^-(\eta_{\rm e})$ at late times 
[cf. Eq.~(\ref{eq:fk-lt})] simplifies to
\begin{equation}
f_\vk^-(\eta_{\rm e})
=\frac{i\, H_0}{2\, \Mp\, \sqrt{{k^3}\,\epsilon_{1-}(\eta_{\rm e})}}.
\label{eq:fk-lt-b0}
\end{equation}
As a consequence, if we can carry out the integral~(\ref{eq:cG40-lk})
describing ${\cal G}_4^0(k)$, then, upon using the above expression for 
$f_\vk^-(\eta_{\rm e})$, we can easily determine the small scale 
behavior of the quantity $k^6\,G_4^0(k)$. 
Recall that, according to the representation~(\ref{eq:rdd-fn}) of the 
delta function that we had considered in the previous section, 
$\epsilon_2^0{}'$ is a constant during the transition 
[cf. Eq.~(\ref{eq:e2p-dt})].
In such a situation, the above integral for ${\cal G}_4^0(k)$ turns out
to be trivial to evaluate and, if we make use of the late time 
modes~(\ref{eq:fk-lt-b0}), we find that we indeed recover the large 
wavenumber behavior~(\ref{eq:G40-mar-lk-fv}) that we had mentioned earlier. 


\subsection{The case with the exponential cut-off}

The main advantage of the approach described above to arrive at the small
scale behavior of the bi-spectrum should be obvious.
We can now make use of the procedure to analytically test the behavior 
of the bi-spectrum at large wavenumbers on the details of the transition. 
With this motivation, let us replace the quantity $\epsilon_2^{0}{}'$ 
by the following alternative representation of the original delta function:
\begin{equation}
\label{eq:transexp}
\epsilon_2^{0}{}'(\eta)
=\frac{6\,\Delta A}{A_+}\; \frac{1}{2\,\varepsilon}\; 
\exp \left(-\frac{\vert\eta-\eta_0\vert}{\varepsilon}\right).
\end{equation} 
If we substitute this expression in Eq.~(\ref{eq:cG40-lk}) describing
${\cal G}_4^0(k)$, it takes the form
\begin{equation}
\label{eq:g4exp}
{\cal G}_4^0(k)
= \frac{3\,H_0\,k}{8\,\Mp^3\,\sqrt{k^3\,\epsilon_1^0}}\;
\frac{6\,\Delta A}{A_+}\;\frac{1}{2\,\varepsilon}\;\,
\int_{\eta_-}^{\eta_+}{\rm d}\eta\, \eta\;
\exp\l(-\frac{\vert \eta -\eta_0\vert}{\varepsilon}
+\,3\,i\,k\,\eta\r),
\end{equation}
where 
\begin{equation}
 \eta_\pm=\eta_0\pm r\;\frac{\varepsilon}{2}, 
\end{equation}
as before, represent the boundaries of the transition. 
However, notice that this definition differs from that of 
Eq.~(\ref{eq:defetapm}). 
Specifically, we have introduced the dimensionless quantity $r$ to 
control the duration of the transition, with $r=1$ leading to the 
case of the step transition that we had considered 
in the last section. 
The justification for the introduction of the additional parameter $r$ 
being that, with the parameterization~(\ref{eq:transexp}), the duration 
of the transition is not necessarily related to the height of 
$\epsilon_2^0{}'$, as it was in the case before. 
In order for Eq.~(\ref{eq:transexp}) to represent the Dirac delta function
faithfully, one should actually choose $\eta_\pm=\pm \infty$, viz. the 
limits wherein $\epsilon_2^0{}'$ vanishes. 
If this condition is not satisfied, there will arise spurious contributions 
proportional to $\epsilon_2^0{}'(\eta_\pm)$.
In what follows, we shall also evaluate these contributions (in order
to highlight a specific point), but they can always be made negligible 
by suitably tuning the parameter $r$. 
The integral~(\ref{eq:g4exp}) can be performed in a straightforward manner,
and we obtain that
\begin{eqnarray}
{\cal G}_4^0(k) 
&=& \f{3\,H_0\,k}{8\,\Mp^3\,\sqrt{k^3\,\epsilon_1^0}}
\biggl\{\f{6\,\Delta A}{A_+}
\l[\f{-1/k_0}{1+9\,k^2\,\varepsilon^2}
+\f{6\,i\,k\,\varepsilon^2}{(1+9\,k^2\,\varepsilon^2)^2}\r]\;
{\rm e}^{-3\,i\,k/k_0}\nn\\ 
& &+\;\epsilon_2^0{}'\l(\eta_\pm\r)\,
\l[\f{\eta_+\,{\rm e}^{3\,i\,k\,\eta_+}}{3\,i\,k-1/\varepsilon}
-\frac{\eta_-\,{\rm e}^{3\,i\,k\,\eta_-}}{3\,i\,k+1/\varepsilon}
+\frac{{\rm e}^{3\,i\,k\,\eta_-}}{\l(3\,i\,k+1/\varepsilon\r)^2}
-\frac{{\rm e}^{3\,i\,k\,\eta_+}}{\l(3\,i\,k-1/\varepsilon\r)^2}
\right]\biggr\}
\end{eqnarray}
and, to arrive at this result, we have assumed that the transition 
is symmetric with $\epsilon_2^0{}'(\eta_+)=\epsilon_2^0{}'(\eta_-)$. 
The above expression is evidently made of two terms, with the second 
one, as already mentioned, originating from the non-zero contributions 
to $\epsilon_2^0{}'$ at the boundaries of the transition. 

\par 

Having arrived at the above expression, we can now evaluate the 
corresponding contribution to the bi-spectrum.
A somewhat lengthy, but straightforward calculation, leads to
\begin{eqnarray}
k^6\, G_4^0(k) 
&=& \frac{27\,\Delta A\, H_0^6}{8\, A_+^2\,\Mp^3\,
\sqrt{2\,\epsilon_{1-}^{3}\l(\eta_{\rm e}\r)}}\; \frac{k}{k_0}\nn\\
& &\times\,
\l[\frac{1}{1+9\,\kappa^2\,k^2/k_0^2}\;\sin\l(\frac{3\,k}{k_0}\r)
+\f{6\,\kappa ^2\,k/k_0}{\l(1+9\,\kappa^2\,k^2/k_0^2\r)^2}\;
\cos\l(\frac{3\,k}{k_0}\r)\r]\nn\\ 
& &-\, \frac{9\, H_0^6\, \kappa\, k/k_0^2}{16\, A_+\,\Mp^3\,
\sqrt{2\,\epsilon_{1-}^{3}\l(\eta_{\rm e}\r)}}\; 
\epsilon_2^0{}'\l(\eta_\pm\r)\;
\Biggl\{\frac{2}{1+9\,\kappa^2\,k^2/k_0^2}\;
\cos\l(\frac{3\, r\,\kappa\,k}{2\,k_0}\r)\;
\sin\l(\frac{3\,k}{k_0}\r)\nn\\
& &-\,\frac{6\,\kappa\, k/k_0}{1+9\,\kappa^2\,k^2/k_0^2}\,
\sin\l(\f{3\,r\,\kappa\,k}{2\,k_0}\r)\, \sin\l(\f{3\,k}{k_0}\r)\nn\\
& &+\,\l[\frac{r\,\kappa }{1+9\,\kappa^2\,k^2/k_0^2}
+ \f{2\,\kappa\,\l(1-9\,\kappa^2\,k^2/k_0^2\r)}{\l(1
+9\,\kappa^2\,k^2/k_0^2\r)^2}\r]\,
\sin\l(\frac{3\,r\,\kappa\,k }{2\,k_0}\r)\,
\cos\l(\frac{3\,k}{k_0}\right)\nn\\
& &+\, \l[\frac{3\,r\,\kappa^2\,k/k_0}{1+9\,\kappa^2\,k^2/k_0^2}
+\frac{12\,\kappa^2\,k/k_0}{\l(1+9\,\kappa^2\,k^2/k_0^2\r)^2}\r]\;
\cos\l(\f{3\,r\,\kappa\,k}{2\,k_0}\r)\, \cos\l(\frac{3\,k}{k_0}\r)\Biggr\},
\label{eq:eco-car}
\end{eqnarray}
where we should remind that, as earlier, $\varepsilon$ has been written 
as $\kappa/k_0$. It is easy to check from the above expression that, in 
the limit corresponding to that of the exact Dirac delta function (i.e. 
as $\varepsilon \to 0$) one recovers the original result that 
$k^6\, G_4^0(k)$ grows linearly as $k/k_0$. Let us now assume that we 
indeed have a faithful representation of the Dirac function so that
$\epsilon_2^0{}'(\eta_\pm)=0$. 
In such a case, we find the leading contribution at large $k$ to be
\begin{eqnarray}
  \lim _{k/k_0\to \infty}
  k^6\, {G}_{4}^0(k)
  &=&\f{3\, \Delta A\, H_0^6}{8\, A_+^2\, \Mp^3\,
    \sqrt{2\,\epsilon_{1-}^{3}\l(\eta_{\rm e}\r)}}\;
    \frac{k_0}{\kappa^2\, k}\;
    \sin\, \l(\f{3\,k}{k_0}\r).\label{eq:eco-ot}
\end{eqnarray}
Note that, importantly, we no longer obtain a plateau in the small scale 
limit, but a term which decreases as $k^{-1}$. This clearly illustrates 
the point that the small scale behavior of the bi-spectrum depends on the 
manner in which the original Dirac delta function and, thereby the 
transition, is smoothened.

\par

\begin{figure}[!t]
\begin{center}
\resizebox{420pt}{280pt}{\includegraphics{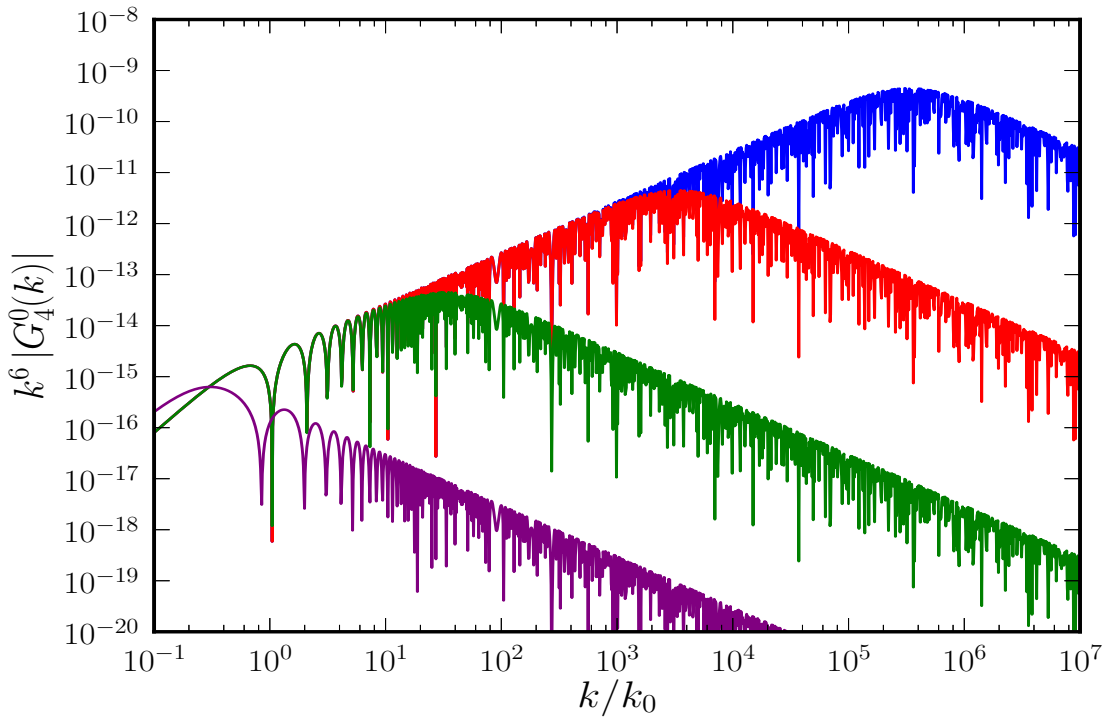}}
\end{center}
\vskip -10pt
\caption{\label{fig:eco}The behavior of the quantity $k^6\, \vert
  G_4^0(k)\vert$ in the Starobinsky model, when the transition has been
  smoothened according to the exponential representation~(\ref{eq:transexp})
  of the original delta function. We have worked with the same set of
  values for the various parameters and the same colors to represent 
  the results corresponding to the different values of $\kappa$, as in 
  the previous figure. The $k^{-1}$ fall-off at large wavenumbers 
  is evident.}
\end{figure}

At this stage, a couple of clarifying remarks are in order regarding the 
result we have obtained above.
If there remains a contribution at the boundary of the transition, that
is to say, if $\epsilon_2^0{}'(\eta_\pm)$ are not exactly zero, we find
that the corresponding contribution to the bi-spectrum at large $k$ is 
given by
\begin{eqnarray}
\lim _{k/k_0\to \infty}
k^6\, {G}_{4}^0(k)
&=& \f{\,3\,H_0^6/k_0}{8\,A_+\, \Mp^3\, 
\sqrt{2\,\epsilon_{1-}^{3}\l(\eta_{\rm e}\r)}}\;
\epsilon_2^0{}'(\eta_\pm)\;
\Biggl[\sin\l(\f{3\,r\,\kappa\,k}{2\,k_0}\r)\, 
\sin\l(\frac{3\,k}{k_0}\r)\nn\\
& &-\,\frac{\kappa\, r}{2}\, \cos\l(\f{3\,r\,\kappa\,k}{2\, k_0}\r)\,
\cos\l(\f{3\,k}{k_0}\right)\Biggr].\quad\label{eq:eco-ac}
\end{eqnarray} 
Since this term depends on the wavenumber only through the trigonometric 
functions, it will lead to a plateau (as in the case of Fig.~\ref{fig:mar}) 
at very small scales, when it begins to dominate the original 
term~(\ref{eq:eco-ot}) which falls as $k^{-1}$.
This will occur at a wavenumber that depends on the overall amplitude 
which, in turn, depends on $r$ and $\epsilon_2^0{}'(\eta_\pm)$.
Let us assume that $\kappa\, r$ is small so that the second term in 
Eq.~(\ref{eq:eco-ac}) above is negligible.
In such a case, upon equating the amplitudes of the first terms in the 
above two equations, we find that the plateau will occur at the wavenumber 
of $k/k_0\simeq \exp\,(r/2)/(3\,\kappa)$.
Upon plotting the complete analytical result~(\ref{eq:eco-car}) for sufficiently 
large wavenumbers, we do observe the plateau, and we also find that the 
plateau indeed begins at wavenumbers corresponding to the above estimate.
But, it ought to be clear that, physically, the plateau should not be present 
since $\epsilon_2^0{}'(\eta_\pm) =0$ is necessary in order to have a faithful 
representation of the Dirac delta function.
In summary, with the original Dirac delta function smoothened and represented 
in terms of an exponential function, we arrive at a behavior wherein the 
contribution to the bi-spectrum due to the transition falls off as $k^{-1}$ at 
large wavenumbers.
In Fig.~\ref{fig:eco}, we have plotted the corresponding results, which 
explicitly illustrate this behavior.


\subsection{Working with a Gaussian representation}

The exponential representation~(\ref{eq:transexp}) of
the Dirac delta function has a cusp at $\eta_0$, and it will be worthwhile 
to extend the above analysis for an even smoother representation.
Towards this end, let us replace the delta function appearing in the
original $\epsilon_2^0{}'(\eta)$ [cf.~Eq.~(\ref{eq:e2d})] with the 
following Gaussian representation: 
\begin{equation}
\epsilon_2^0{}'(\eta) =\f{6\,\Delta A}{\sqrt{\pi}\, A_+\, \varepsilon}\;
{\rm exp}\left[-\frac{\l(\eta-\eta_0\r)^2}{\varepsilon^2}\right].\label{eq:gr}
\end{equation}
In such a case, upon carrying out the integral~(\ref{eq:cG40-lk}) from
$\eta_-=\eta_0-r\,\varepsilon/2$ to $\eta_+=\eta_0+r\,\varepsilon/2$, and
ignoring the contributions due to the end points of the transition (for
reasons discussed in the previous subsection), we obtain that
\begin{eqnarray}
{\cal G}_4^0(k) 
&=& \f{3\, H_0\, k}{8\, \Mpl^3\, \sqrt{k^3\,\epsilon_1^0}}\,
\f{3\, \Delta A}{A_+}\, 
\l(\f{3\, i\, k\,\varepsilon^2}{2}-\f{1}{k_0}\r)\;
{\rm e}^{-9\,k^2\, \varepsilon^2/4}\;\,
{\rm e}^{-3\, i\, k/k_0}\nn\\
& &\times\,\biggl[{\rm erf}\l(\frac{r}{2}-\frac{3\,i\,k\,\varepsilon}{2}\r)-
{\rm erf}\l(-\frac{r}{2}-\frac{3\,i\,k\,\varepsilon}{2}\r)\biggr],
\end{eqnarray}
where ${\rm erf}(z)$ denotes the error 
function~\cite{Gradshteyn:1965aa,Abramovitz:1970aa}.
One can then obtain the corresponding contribution to the bi-spectrum 
to be
\begin{eqnarray}
k^6\, G_4^0(k) 
&=& \frac{27\,\Delta A\, H_0^6}{32\, A_+^2\,\Mp^3\,
\sqrt{2\,\epsilon_{1-}^{3}\l(\eta_{\rm e}\r)}}\;
{\rm e}^{-9\,k^2\,\varepsilon^2/4}\nn\\
& &\times\,\Biggl\{\biggl[
{\rm erf}\l(\frac{r}{2}-\frac{3\,i\,k\,\varepsilon}{2}\r)-
{\rm erf}\l(-\frac{r}{2}-\frac{3\,i\,k\,\varepsilon}{2}\r)\biggr]\,
\l(\f{3\, k^2\,\varepsilon^2}{2}+\f{i\,k}{k_0}\r)\;
{\rm e}^{-3\, i\, k/k_0}\nn\\
& &+\,\biggl[{\rm erf}\l(\frac{r}{2}+\frac{3\,i\,k\,\varepsilon}{2}\r)-
{\rm erf}\l(-\frac{r}{2}+\frac{3\,i\,k\,\varepsilon}{2}\r)\biggr]\,
\l(\f{3\, k^2\,\varepsilon^2}{2}-\f{i\,k}{k_0}\r)\;
{\rm e}^{3\, i\, k/k_0}\Biggl\}.\nn\\
\end{eqnarray}
If we now assume $r$ to be sufficiently large, then we find that this
expression simplifies to
\begin{eqnarray}
k^6\, G_4^0(k) 
&=& \frac{27\,\Delta A\, H_0^6}{8\, A_+^2\,\Mp^3\,
\sqrt{2\,\epsilon_{1-}^{3}\l(\eta_{\rm e}\r)}}\;\,
{\rm e}^{-9\,k^2\,\varepsilon^2/4}
\nn\\ & &\times\,
\l[\frac{k}{k_0}\,\sin\l(\frac{3\,k}{k_0}\r)
+\frac{3\,k^2\,\varepsilon^2}{2}\,
\cos\l(\frac{3\,k}{k_0}\r)\r],
\end{eqnarray}
which reduces to the original result [viz. Eq.~(\ref{eq:G40-oar-lk})] 
involving the sharp step in the limit $\varepsilon\to 0$.
The reason for assuming $r$ to be sufficiently large is the same 
as the reason we had attributed in the case of the exponential 
representation discussed in the previous sub-section.
If $\epsilon_2^0{}'(\eta_\pm)$ do not vanish, then the 
Gaussian~(\ref{eq:gr}) ceases to be a faithful representation of 
the original delta function, and it can then lead to incorrect
contributions. 
In fact, it can be easily established analytically that the immediate 
sub-leading term (for a finite $r$) contains an additional Gaussian 
growth and, when taken into account along with the overall Gaussian 
suppression encountered above, it leads to a spurious plateau at 
large wavenumbers, just as in the exponential case.
It should now be evident from the examples we have considered that the 
smoother $\epsilon_2'(\eta)$ is during the transition, the sharper 
is the cut-off in the scalar bi-spectrum at large wavenumbers.

\par

More generally, it should be clear from the above discussion that, 
the contribution to the bi-spectrum due to the transition contains 
a cut-off at large wavenumbers for an arbitrary smooth transition. 
The specific case that we had considered in the last section wherein 
the form of smoothening had led to a plateau therefore appears to be 
a very particular situation. 
The exact nature of the cut-off, of course, depends on the precise 
form of the transition (and is therefore not necessarily proportional 
to $k^{-1}$ or suppressed exponentially), but our analysis unambiguously 
shows that a cut-off is generically present. 
As we shall illustrate in the next section, these conclusions are also
corroborated by numerical calculations.


\section{Comparison with the numerical results from BINGO}
\label{sec:cnr}

Recently, we have developed an efficient and accurate numerical code,
called BINGO, to evaluate the scalar bi-spectrum in inflationary models 
involving the canonical scalar field~\cite{Hazra:2012yn}. In this section,
we shall make use of BINGO to numerically investigate the effects of 
smoothening the sharp transition in the Starobinsky model. 
In place of the actual potential~(\ref{eq:p-sm}), we shall work with the 
following potentials that have been smoothened in two different fashion:
\begin{eqnarray}
V_1(\phi) 
&=& V_0+\f{1}{2}\, (A_++A_-)\, (\phi - \phi_0)
+ \f{1}{2}\, (A_+-A_-)\, (\phi - \phi_0)\;
{\rm tanh}\, \l(\f{\phi - \phi_0}{\Delta \phi}\r),\label{eq:ssm1}\\
V_2(\phi) 
&=& V_0+\f{1}{2}\, (A_++A_-)\, (\phi - \phi_0)
+ \f{1}{2}\, (A_+-A_-)\,\Delta \phi\;\;
{\rm ln}\biggl[{\rm cosh}\,\l(\f{\phi - \phi_0}{\Delta \phi}\r)\biggr],
\label{eq:ssm2}
\end{eqnarray}
both of which, as is required, reduce to the shape of the original
potential in the limit $\Delta\phi\to0$.
Also, it should be clear that, in the above potentials, instead of the 
original, infinitely sharp transition, the field will make the 
transition over the width $\Delta\phi$ in field space.  

\par

\begin{figure}[!t]
\vskip -22pt
\resizebox{405pt}{270pt}{\includegraphics{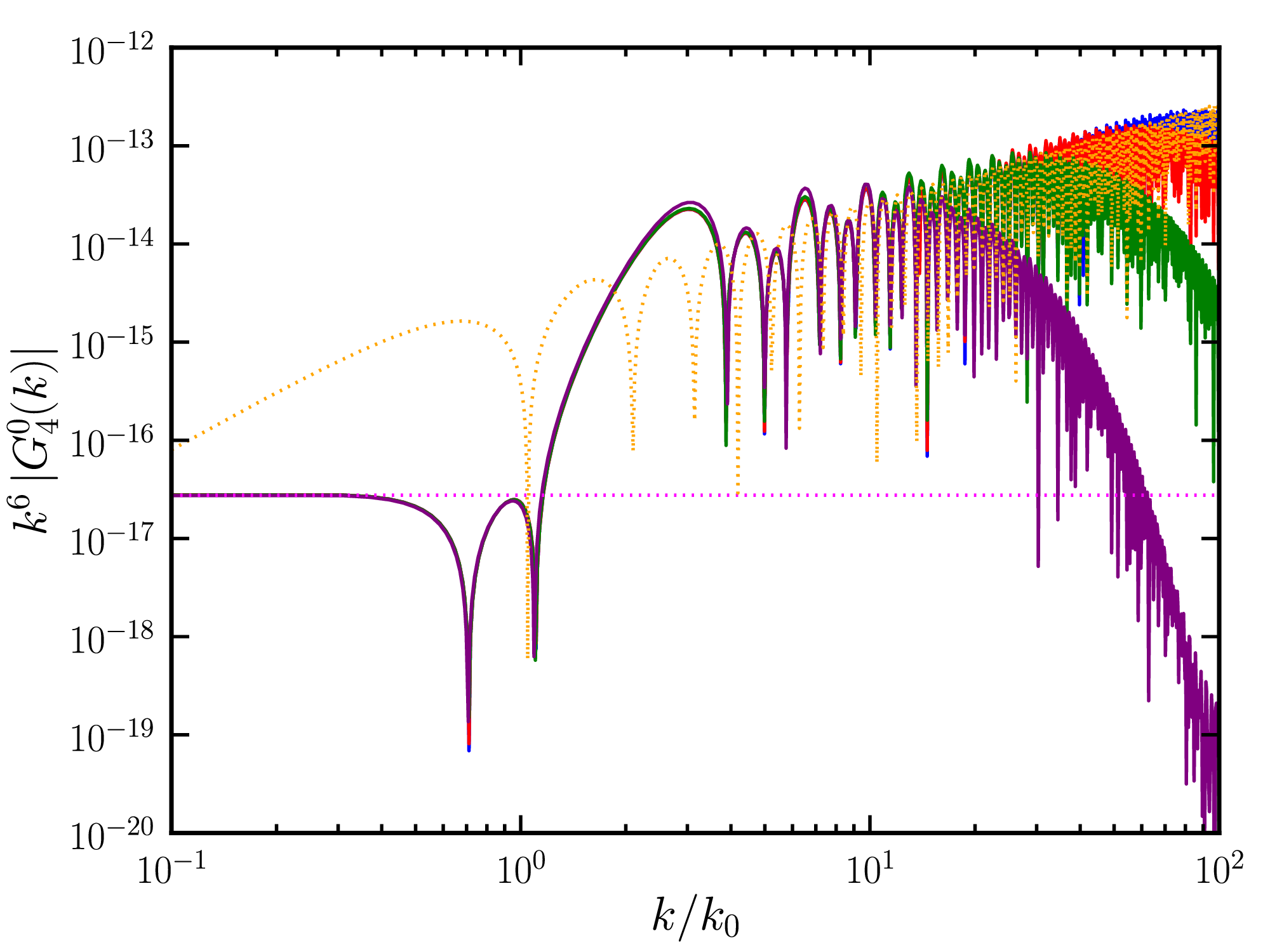}}
\vskip -5pt
\resizebox{405pt}{270pt}{\includegraphics{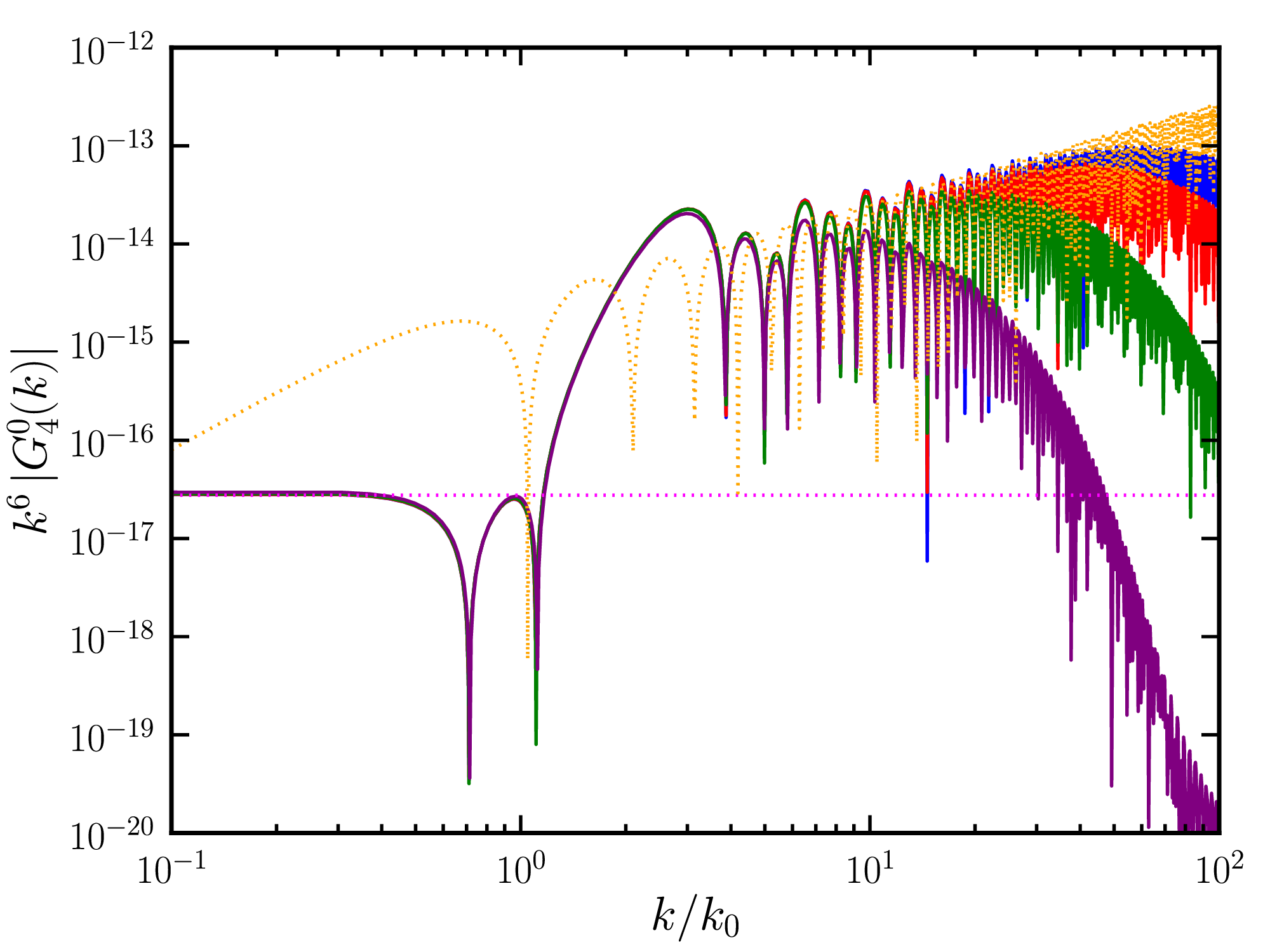}}
\vskip -5pt
\caption{\label{fig:nr}The behavior of the quantity $k^6\, \vert G_4^0(k)
\vert$ with a smoothened Starobinsky model that is described by the 
potentials~(\ref{eq:ssm1}) (on top) and~(\ref{eq:ssm2}) (below).
These results have been obtained using BINGO, which is a recently developed 
numerical code to evaluate the scalar bi-spectrum~\cite{Hazra:2012yn}. 
We have worked with the same values for the parameters $V_0$, $A_+$ and 
$A_-$ as in the earlier figures, but have varied $\Delta \phi$.
The plots correspond to $\Delta\phi/\phi_0$ of $1/7500$ (in blue), $1/5000$ 
(in red), $1/2500$ (in green) and $1/1000$ (in purple).  
As in Fig.~\ref{fig:oar}, we have also plotted the asymptotic behavior 
of the analytical result from the original, unsmoothened, Starobinsky model
in the limit of small (in magenta) and large (in orange) wavenumbers.
It should be clear from the two figures that, as the step is smoothened or, 
equivalently, the transition is widened, the growth due to term involving 
$V_{\phi\phi}$ is considerably reduced and its contribution ceases to be 
important at suitably large $k$.}
\end{figure}

While the details of the numerical procedures to compute the scalar 
bi-spectrum can be found in our earlier work~\cite{Hazra:2012yn}, we
believe that a couple of brief and generic remarks are in order at 
this stage of our discussion.
Given a potential and the value of the parameters that describe it, 
the background evolution is completely determined by the initial 
conditions on the scalar field.
If one further assumes that the perturbations are in the Bunch-Davies 
vacuum at sufficiently early times, the quantities that characterize 
the perturbations---such as the power spectrum and the 
bi-spectrum---are uniquely determined as well.
In order to compare with the analytical results we have obtained in
the previous section, using BINGO, we numerically compute the 
contribution to the fourth term of the bi-spectrum, viz. $G_4(k)$, 
assuming that the quantity ${\dot \epsilon}_2$ [cf. Eq.~(\ref{eq:e2d})] 
is determined {\it only}\/ by the term involving $V_{\phi\phi}$ 
corresponding to the smoothened potentials~(\ref{eq:ssm1}) 
and~(\ref{eq:ssm2}).
We work with the same values of the original parameters $V_0$, $A_+$ and  
$A_-$ that we had considered in the previous three figures, but vary 
$\Delta\phi$ over a suitable range.
In Fig.~\ref{fig:nr}, we have plotted the resulting contribution to the 
bi-spectrum for a few different values of $\Delta\phi$.
It should be clear from the figure that the smoother the transition 
the more stunted is the growth at large wavenumbers. 

\par

At this stage, it is important that we highlight the results we have obtained 
and also discuss earlier efforts in similar situations.
As we had outlined in the introductory section, our main aim had been to 
illustrate that an indefinite growth in the bi-spectrum is unphysical 
and that it is related to the infinitely sharp transition that one encounters 
in the original Starobinsky model.
Moreover, we had intended to show that, for any finite and smooth transition, 
the scale invariance of the bi-spectrum will be restored at suitably large
wavenumbers. 
Evidently, we have been able to establish these two points both analytically 
and numerically.
While it seems natural to expect that the indefinite growth will be suppressed  
if the transition is smoothened~\cite{Arroja:2012ae}, it had not been established
earlier. 
We have been able to explicitly show that this is indeed the case.
However, it should be clear from our discussion in the last two sections that,
whereas the contribution due to a smooth transition is generically suppressed 
at large wavenumbers, the details of the suppression depends on the way in
which the discontinuity is smoothened.

\par

In fact, we should also emphasize here that there has also been prior efforts in 
studying the bi-spectrum in similar scenarios.
These previous efforts had made use of the so-called generalized slow roll approximation, 
which is a generic analytical method to study cases involving short periods of 
fast roll (see, for instance, Refs.~\cite{Dvorkin:2009ne,Hu:2011vr}).
Using the approach, it
\clearpage
\noindent
has been shown that sufficiently sharp steps in the potential 
can lead to a strong burst of oscillations before the bi-spectrum turns scale invariant 
at small scales (in this context, see Refs.~\cite{Adshead:2011jq,Adshead:2013zfa}).   
There exist some similarities and certain differences between our work and the
earlier efforts based on the generalized slow roll approximation.
Evidently, the results from the generalized slow roll approach can be expected to be
broadly applicable to the linear potential of our interest here and, in this sense, 
our effort can be considered to be similar to the prior efforts, but without the
constant term $V_0$.
However, the presence of the constant $V_0$ in the potential (which we have assumed 
to be the dominant term) turns out to be important in our approach as it ensures that 
the background evolution is rather close to that of de Sitter.
Moreover, due to this reason (and also because of the linear nature of the potential), 
the de Sitter modes prove to be a very good approximation to the scalar 
perturbations\footnote{It is well known that a de Sitter background and the de Sitter
solutions for the scalar modes are generally a good approximation in most slow roll 
scenarios.
However, in the Starobinsky model, one finds that, since $V_0$ is the dominant term and 
the fact that the potential is linear make them particularly good approximations.
A large $V_0$ will, evidently, ensure that the background behaves essentially as that of 
de Sitter. 
Before the transition, since $\epsilon_1$ is small and constant, $\epsilon_2\simeq 4\, 
\epsilon_1$  is small and constant as well, while $\epsilon_3$ vanishes. 
Interestingly, it can be shown that, after the transition, as $\epsilon_1$ continues to 
remain small (because of the dominant $V_0$ term) and the potential is linear, certain 
cancellations occur in the expression for the quantity $z''/z$ as a result of which the 
scalar modes are described very well by the de Sitter solutions~\cite{Martin:2011sn}.}.
Further, while the bi-spectrum in the earlier efforts was evaluated up to the first order
in the generalized slow roll approximation, such a limitation does not apply to our
approach.
We believe that these three points make the approximations we have adopted work well, 
as is confirmed by the numerical analysis. 
But, our approach is specifically designed for the case of the linear potential, 
dominated by a constant term.
In contrast, the generalized slow roll approach can be applied to a wider class of 
potentials.
We feel it will be interesting to examine if the linear growth in the original 
Starobinsky model can be reproduced, say, at a certain order, in the generalized 
slow roll approach.
Nevertheless, we should stress the fact that the restoration of scale invariance in 
the scenarios studied using the generalized slow roll 
approximation~\cite{Adshead:2011jq,Adshead:2013zfa} corroborate the main conclusions 
that we have arrived at here.


\section{Discussion}\label{sec:d}

It is well known that periods of departure from slow roll will lead to
certain features in the inflationary scalar power spectrum corresponding 
to modes that leave the Hubble radius during the epochs of fast roll.
However, in the case of the power spectrum, scale invariance is always
restored when slow roll has been reestablished (see, for example, Refs.~\cite{Ashoorioon:2006wc,Ashoorioon:2008qr,Hazra:2010ve}).  Such a behavior
occurs independent of the sharpness or the extent of the deviation
from slow roll. In complete contrast to the behavior of the power
spectrum, in the case of the Starobinsky model, it has been found that
the abrupt transition that occurs leads to a term which grows linearly
at large wavenumbers in the scalar bi-spectrum.  Importantly, this
occurs despite the fact that slow roll is restored a little while
after the field crosses the discontinuity in the Starobinsky model.

\par

Clearly, the continuing growth of the scalar bi-spectrum in the
Starobinsky model is an artifact of the infinitely sharp transition,
and one would imagine that the bi-spectrum will turn scale invariant
at sufficiently small scales if the transition is made smoother.
In this work, we have shown, both analytically and numerically, that 
this expected behavior indeed
occurs. Analytically, we have been able to show that a 
sufficient smoothening of the transition leads to a truncation of the growth and an eventual sharp fall-off at a suitably 
large wavenumber which is related to the width of the transition. 
Numerically, we find that, as the sharpness of the transition is 
decreased, the width of the feature that arises as a result reduces, 
with the contribution due to the transition ceasing to be important at 
large wavenumbers.
In such situations, the various remaining contributions to the bi-spectrum
that we have calculated in some detail earlier (see Ref.~\cite{Martin:2011sn})
become important.

\par

We should point out here that the restoration of scale invariance of
the bi-spectrum at large wavenumbers can occur in two different ways. 
Evidently, the complete bi-spectrum is the sum of the contribution 
due to the $V_{\phi \phi}$ term in $\epsilon_2'$ (which had been
the focus of our attention here) and the contributions due to all 
the other terms (that have been calculated earlier in 
Ref.~\cite{Martin:2011sn}). 
The amplitude of the remaining contributions goes to a constant value 
at small scales~\cite{Martin:2011sn}.
Hence, scale invariance of the complete bi-spectrum is restored if the 
contribution due to the $V_{\phi \phi}$ term either itself turns a 
constant or vanishes at large wavenumbers. 
Our analysis suggests that both cases can arise depending on the manner
in which the transition is actually smoothened, i.e. it depends on the 
microphysics of the transition. 
We should emphasize that a generic smoothening of the transition does 
not necessarily lead to an exponential cut-off in the contribution due 
to the $V_{\phi \phi}$ term, as one might naively be tempted to deduce, 
possibly guided by the example of particle production presented in the 
introductory section. 
But, such a behavior is nevertheless consistent with the scale invariance 
of the total bi-spectrum at small scales.
This is because of the reason that there exist different ways to arrive 
at a scale invariant behavior at large $k$, as we have explained above.

\par

Before concluding, it is worthwhile that we touch upon two related
points.  Firstly, it would be interesting to examine if there can 
exist conditions under which the power spectrum itself may exhibit 
the behavior as the scalar bi-spectrum does in the case of the 
Starobinsky model, i.e. grow indefinitely at large wavenumbers. The 
second point that is worth considering concerns the implication of 
such behavior for the higher order correlation functions such as the 
tri-spectrum. Let us now turn to discuss these two points.

\par 

The fact that the discontinuity in the second derivative of the
potential leads to the growth in the bi-spectrum suggests that 
one can expect such a discontinuity in the first derivative of 
the potential to influence the power spectrum. However, a 
discontinuity in the first derivative of the potential  would
imply that the first slow roll parameter itself would grow large, 
thereby even ending inflation. In such a case, two possibilities 
can arise. Either, inflation is completely terminated, never to be 
restored. Or, the departure from the accelerated expansion may be 
of an extremely short duration and the shape of the potential 
permits inflation to restart. The former situation does not help, 
whereas the latter scenario, dubbed punctuated
inflation~\cite{Jain:2008dw,Jain:2009pm}, is indeed a genuine
possibility. However, inflation reestablished in such situations 
proves to be of the slow roll type, which also restores scale 
invariance of the power spectrum.

\par 

One can expect that the tri-spectrum will involve one higher order
slow parameter beyond the third. If so, then the tri-spectrum can, 
in fact, be expected to diverge in the case of the Starobinsky model, 
since the fourth slow roll parameter $\epsilon_4$ would. Actually, 
not only the tri-spectrum, this conclusion may apply to all the higher 
order correlations functions as well. For any transition of a finite 
width, one can expect the tri-spectrum and the higher order correlation
functions to exhibit a rather sharp rise for modes that leave the
Hubble scale during the transition. These issues are worth studying
closer.


\section*{Acknowledgments}

LS wishes to thank Institut d'Astrophysique de Paris, France, for
support and hospitality during a visit, when this work was largely
carried out. 
JM would like to thank the Indian Institute of Technology Madras,
Chennai, India, for support and warm hospitality during a visit where 
this work was completed.
DKH wishes to acknowledge support from the Korea Ministry of Education, 
Science and Technology, Gyeongsangbuk-Do and Pohang City for Independent 
Junior Research Groups at the Asia Pacific Center for Theoretical Physics,
Pohang, Korea.


\bibliographystyle{JHEP}
\bibliography{ng}


\end{document}